 \documentclass[twocolumn,tighten,times]{aastex61}

\definecolor{purple}{rgb}{0.4,0.0,1.} 
\definecolor{lightgreen}{rgb}{0.67,0.87,0.0}

\accepted{by The Astrophysical Journal on 29th April, 2018}
\shorttitle{NGFS:~IV.~Mass and Age Bimodality of Nuclear Clusters in the Fornax Core Region}
\shortauthors{Y. Ordenes-Brice\~no et al.}
\begin{document}
\title{The Next Generation Fornax Survey (NGFS):~IV.~Mass and Age Bimodality of Nuclear Clusters in the Fornax Core Region}

\correspondingauthor{Yasna Ordenes-Brice\~no}
\email{yordenes@astro.puc.cl}

\author[0000-0001-7966-7606]{Yasna Ordenes-Brice\~no}
\altaffiliation{PUC-HD Graduate Student Exchange Fellow}
\affiliation{Institute of Astrophysics, Pontificia Universidad Cat\'olica de Chile, Av.~Vicu\~na Mackenna 4860, 7820436 Macul, Santiago, Chile}
\affiliation{Astronomisches Rechen-Institut, Zentrum f\"ur Astronomie der Universit\"at Heidelberg, M\"onchhofstra{\ss}e 12-14, D-69120 Heidelberg, Germany}

\author[0000-0003-0350-7061]{Thomas H.~Puzia}
\affiliation{Institute of Astrophysics, Pontificia Universidad Cat\'olica de Chile, Av.~Vicu\~na Mackenna 4860, 7820436 Macul, Santiago, Chile}

\author[0000-0001-8654-0101]{Paul~Eigenthaler}
\altaffiliation{CASSACA Postdoctoral Fellow}
\affiliation{Institute of Astrophysics, Pontificia Universidad Cat\'olica de Chile, Av.~Vicu\~na Mackenna 4860, 7820436 Macul, Santiago, Chile}

\author[0000-0003-3009-4928]{Matthew A.~Taylor}
\altaffiliation{Gemini Science Fellow}
\affiliation{Gemini Observatory, Northern Operations Center, 670 North A'ohoku Place, Hilo, HI 96720, USA}

\author[0000-0003-1743-0456]{Roberto~P.~Mu\~noz}
\affiliation{Institute of Astrophysics, Pontificia Universidad Cat\'olica de Chile, Av.~Vicu\~na Mackenna 4860, 7820436 Macul, Santiago, Chile}

\author[0000-0003-1632-2541]{Hongxin Zhang}
\altaffiliation{FONDECYT Postdoctoral Fellow}
\affiliation{Institute of Astrophysics, Pontificia Universidad Cat\'olica de Chile, Av.~Vicu\~na Mackenna 4860, 7820436 Macul, Santiago, Chile}

\author[0000-0002-5897-7813]{Karla~Alamo-Mart\'inez}
\altaffiliation{FONDECYT Postdoctoral Fellow}
\affiliation{Institute of Astrophysics, Pontificia Universidad Cat\'olica de Chile, Av.~Vicu\~na Mackenna 4860, 7820436 Macul, Santiago, Chile}

\author[0000-0002-3004-4317]{Karen X.~Ribbeck}
\affiliation{Institute of Astrophysics, Pontificia Universidad Cat\'olica de Chile, Av.~Vicu\~na Mackenna 4860, 7820436 Macul, Santiago, Chile}

\author[0000-0002-1891-3794]{Eva K.\ Grebel}
\affiliation{Astronomisches Rechen-Institut, Zentrum f\"ur Astronomie der Universit\"at Heidelberg, M\"onchhofstra{\ss}e 12-14, D-69120 Heidelberg, Germany}

\author[0000-0002-5322-9189]{Sim\'on~\'Angel}
\affiliation{Institute of Astrophysics, Pontificia Universidad Cat\'olica de Chile, Av.~Vicu\~na Mackenna 4860, 7820436 Macul, Santiago, Chile}

\author[0000-0003-1184-8114]{Patrick C{\^o}t{\'e}}
\affiliation{NRC Herzberg Astronomy and Astrophysics, 5071 West Saanich Road, Victoria, BC V9E 2E7, Canada}

\author[0000-0002-8224-1128]{Laura Ferrarese}
\affiliation{NRC Herzberg Astronomy and Astrophysics, 5071 West Saanich Road, Victoria, BC V9E 2E7, Canada}

\author[0000-0002-2363-5522]{Michael Hilker}
\affiliation{European Southern Observatory, Karl-Schwarzchild-Str. 2, D-85748 Garching, Germany}

\author[0000-0002-7214-8296]{Ariane~Lan\c{c}on}
\affiliation{Observatoire astronomique de Strasbourg, Universit\'e de Strasbourg, CNRS, UMR 7550, 11 rue de l'Universite, F-67000 Strasbourg, France}

\author[0000-0003-4197-4621]{Steffen Mieske}
\affiliation{European Southern Observatory, 3107 Alonso de C\'ordova, Vitacura, Santiago}

\author[0000-0002-5665-376X]{Bryan W.~Miller}
\affiliation{Gemini Observatory, South Operations Center, Casilla 603, La Serena, Chile}

\author[0000-0002-2204-6558]{Yu Rong}
\altaffiliation{CASSACA Postdoctoral Fellow}
\affiliation{Institute of Astrophysics, Pontificia Universidad Cat\'olica de Chile, Av.~Vicu\~na Mackenna 4860, 7820436 Macul, Santiago, Chile}

\author[0000-0003-4945-0056]{Ruben S\'anchez-Janssen}
\affiliation{STFC UK Astronomy Technology Centre, Royal Observatory, Blackford Hill, Edinburgh, EH9 3HJ, UK}

\begin{abstract}
We present the analysis of 61 nucleated dwarf galaxies in the central regions ($\la\!R_{\rm vir}/4$) of the Fornax galaxy cluster.~The galaxies and their nuclei are studied as part of the {\it Next Generation Fornax Survey} (NGFS) using optical imaging obtained with the Dark Energy Camera (DECam) mounted at Blanco/CTIO and near-infrared data obtained with VIRCam at VISTA/ESO.~We decompose the nucleated dwarfs in nucleus and spheroid, after subtracting the surface brightness profile of the spheroid component and studying the nucleus using PSF photometry.~In general, nuclei are consistent with colors of confirmed metal-poor globular clusters, but with significantly smaller dispersion than other confirmed compact stellar systems in Fornax.~We find a bimodal nucleus mass distribution with peaks located at $\log({\cal M_*}/M_\odot)\!\simeq\!5.4$ and $\sim\,6.3$.~These two nucleus sub-populations have different stellar population properties, the more massive nuclei are older than $\sim\!2$\,Gyr and have metal-poor stellar populations ($Z\leq0.02\,Z_\odot$), while the less massive nuclei are younger than $\sim\!2$\,Gyr with metallicities in the range $0.02\!<\!Z/Z_\odot\!\leq\!1$.~We find that the nucleus mass (${\cal M}_{\rm nuc}$)\,vs.\,galaxy mass (${\cal M}_{\rm gal}$) relation becomes shallower for less massive galaxies starting around $10^8\,M_\odot$ and the mass ratio $\eta_n\!=\!{\cal M}_{\rm nuc}/{\cal M}_{\rm gal}$ shows a clear anti-correlation with ${\cal M}_{\rm gal}$ for the lowest masses, reaching $10\%$.~We test current theoretical models of nuclear cluster formation and find that they cannot fully reproduce the observed trends.~A likely mixture of {\it in-situ} star formation and star-cluster mergers seem to be acting during nucleus growth over cosmic time.
\end{abstract}


%%%%%%%%%%%%%%%%%%%%%%%%%%%%%%%%%%%%%%%%%%%%%%%%%%%%%%%%%%%%

\section{Introduction} \label{sec:intro}
Dwarf galaxies dominate the galaxy number density in dense environments.~Whether they contain a compact stellar nucleus at their centers, or not, is an important distinction among the dwarf galaxy population.~Nuclear clusters are very dense stellar systems with sizes similar to globular cluster (GCs, $\sim\!3-10$\,pc) \citep{Boker04, Cote06, Turner12, DenBrok14, Georgiev14, puz14} but with a broader range of masses $10^5\!-\!10^8\,M_\odot$ \citep[e.g.,][]{Walcher06, Georgiev16, Spengler17}.~Nuclei are a common characteristic in galaxies, from dwarfs to giants.~The nucleation fraction can reach around 70-80$\%$ and is independent of the galaxy morphology \citep{Boker04, Cote06, Georgiev09, Turner12, DenBrok14, Georgiev14, Munoz15, eigenthaler2018}.~However, the nucleation fraction decreases with luminosity \citep[e.g.,][]{Munoz15}, going from as high as $\sim\!90\%$ for galaxies brighter than $M_i\!\leq\!-16$\,mag to $0\%$ for $M_i\!\leq\!-10$\,mag.~This may be related to instrument sensitivity limits, beyond which it becomes harder to detect the lowest surface brightness spheroids and, thus, to associate a nucleus with a low-surface-brightness galaxy spheroid.~Towards the bright galaxy regime, it has been noticed that nuclei are no longer detected for galaxies with $M_B\!<\!-19.5$\,mag.~This might be related to the fact that the central parsecs of bright galaxies can have complex surface brightness profiles, which makes it difficult to separate the galaxy light from the nucleus light, if at all present \citep[e.g.][]{Cote06, Turner12}.~Another reason is that central super massive black holes (SMBHs) co-existing with nuclear clusters can dissolve the central cluster if the SMBH is massive enough, increasing its sphere of influence to radii similar to those of the nuclear cluster \citep[e.g.][]{antonini13}.

Nuclear cluster studies have revealed several correlations between nuclei and their host galaxy, such as the nucleus to galaxy mass relation, their velocity dispersion and galaxy mass\footnote{For late-type galaxies, this would be the bulge mass.}, and the size-luminosity relation \citep{Boker04, Turner12, Georgiev14, DenBrok14, Spengler17}.~These correlations indicate a connection between the nucleus and the formation of its parent galaxy.~Furthermore, the stellar population properties of nuclei seem to be more complex than first thought, revealing multiple stellar populations rather than being old and metal-poor objects.~Using spectra, \cite{Rossa06} found in a sample of 40 late-type galaxies (LTGs) that the luminosity weighted ages of half of the nuclei is younger than 1\,Gyr, within a range from 10\,Myr to very old ages $\ga\!10$\,Gyr \citep[see also][]{Walcher06}.~For 26 early-type galaxies (ETGs) in the Virgo cluster, \cite{Paudel11} found spectroscopic evidence that the age distribution of nuclei is dominated by young ages.~In terms of metallicity, their work revealed a broad metallicity distribution of the nuclei, from $-1.22$\,dex to $+0.18$\,dex, which was wider compared to the host galaxy metallicity range.~When analyzing age and metallicity distributions in radial profiles using bins along the major axis of the dwarf galaxies, \citeauthor{Paudel11}~found that the inner bins are dominated by young ages and broader metallicity distributions than outer regions.~\cite{Spengler17} observed in Virgo cluster galaxies that, on average, the nuclei and host galaxies have similar metallicities with a mean metallicity of $0.07\pm0.3$\,dex, but if they exclude the galaxies that deviate from the mass-metallicity relation then nuclei are on average $0.20$\,dex more metal-rich than their host galaxies.~Clearly, conducting deep, homogeneous, panchromatic nuclear cluster searches, based on wide-field imaging data will allow us to focus on the faint {\it and} bright nucleated galaxy regime at the same time and may help put some of these seemingly contradicting observations in context.

Considerable observational efforts were recently undertaken to map the nearby galaxy cluster regions in Virgo and Fornax, with deep, homogeneous, wide-field surveys reaching low surface-brightness dwarf galaxies \citep[$\langle\mu_{e,i}\rangle\!\simeq\!29$ mag arcsec$^{-2}$, see e.g.][]{fer12, mih05, mih17, Munoz15, eigenthaler2018}.~This is the case for the {\it Next Generation Fornax Survey} (NGFS) that covers a large area of the Fornax galaxy cluster with optical and near-infrared observations.~The NGFS team has identified 258 dwarf galaxy candidates in the central Fornax regions ($R_{\rm \small{NGC1399}}\leq 350$\,kpc).~From the total sample (258) only 75 dwarfs are nucleated (29\%) \citep{eigenthaler2018}.~So far, the faintest nucleated dwarf candidate in the NGFS sample has an absolute magnitude of $M_i\!\simeq\!-10.8$~mag, indicating that the luminosity ratio between the nucleus and the host spheroid  $\eta_{L}\!=\!{\cal L}_{\rm nucleus}/{\cal L}_{\rm spheroid}$ is significantly higher for the faint galaxies than previously thought.~Earlier studies of bright ETGs found $\langle\eta_L\rangle\!=\!0.41\%$ in Fornax \citep{Turner12} and $\langle\eta_L\rangle\!=\!0.3\%$ for Virgo \citep{Cote06}.~The NGFS sample pushes the study of galaxy nuclei into the faint luminosity regime.~In dense galaxy cluster environments, nucleated galaxies have been studied up to now in galaxies brighter than $M_i\!\simeq\!-15$\,mag in Fornax and Virgo \citep{Turner12, Cote06} using the Advance Camera Survey (ACS) in the Hubble Space Telescope (HST), with a sample of 31 and 45 nuclei, respectively.

\begin{figure*}[t!]
    %trim=left bottom right top
     \centering
     \includegraphics[width=0.95\textwidth]{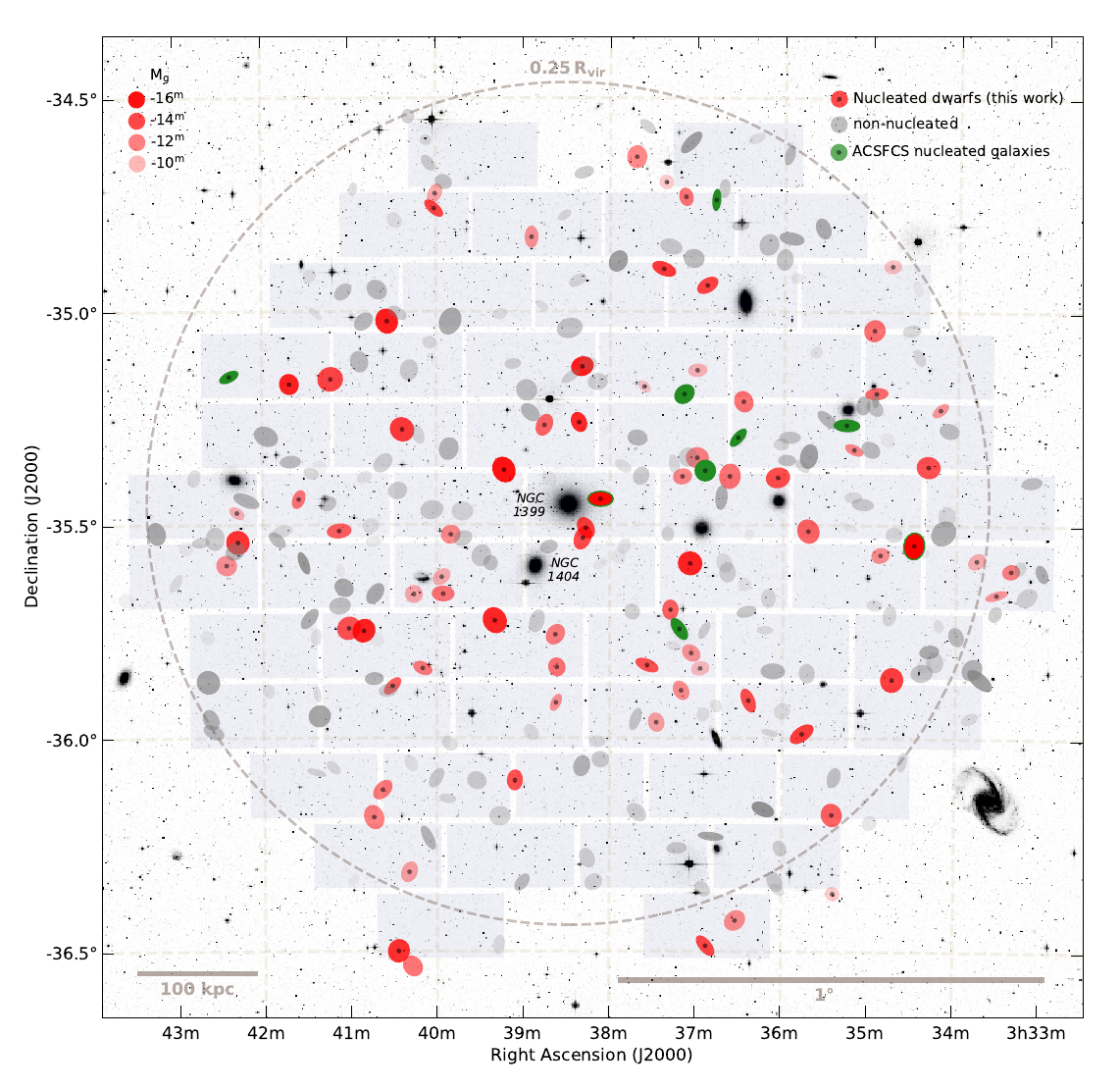}
     \caption{Central region of the Fornax galaxy cluster of the inner $\sim\!25\%$ of the virial radius ($R_{\rm vir}/4$, see dashed circle), showing the spatial distribution of nucleated dwarfs (red symbols), non-nucleated dwarfs (grey symbols), and the nucleated galaxies studied by the ACSFCS \citep[green symbols; see][]{Cote06}.~The field is centered on the giant elliptical cD galaxy NGC\,1399, which is located at the center of the field.~The symbol transparency parameterizes the galaxy luminosity as indicated in the top left corner, i.e.~more transparent symbols indicate fainter galaxies, while the ellipticity and position angle of the symbol represent the same parameters of the corresponding  galaxy spheroid.~Two red ellipses with green edges mark the two dE,N galaxies (FCC\,202 and FCC\,136) that are included in both the ACSFCS sample and this work.}
     \label{fig:spatial}
\end{figure*}

In this work we explore, for the first time, the faint nucleated dwarf galaxy luminosity regime in terms of their photometric properties and scaling relations.~We assume a distance modulus of $31.51\pm0.03$\,mag for Fornax, which corresponds to a distance of $20$\,Mpc \citep{bla09}.~The derived magnitudes in optical passbands are all in the AB system and the NIR magnitudes where transformed from Vega to the AB system using $\Delta m_{K_s}\!=\!1.85$\,mag and $\Delta m_{J}\!=\!0.91$\,mag \citep{Blan07}.

%%%%%%%%%%%%%%%%%%%%%%%%%%%%%%%%%%%%%%%%%%%%%%%%%%%%%%%%%%%%
\section{Observations}
The {\it Next Generation Fornax Survey} (NGFS) is an ongoing deep multi-wavelength survey that covers the entire Fornax galaxy cluster out to its virial radius (1.5\,Mpc).~NGFS uses the Cerro Tololo Inter-American Observatory (CTIO) 4-meter Blanco telescope in combination with the Dark Energy Camera \citep[DECam][]{Flaugher15} for optical photometry, as well as the European Southern Observatory (ESO) 3.7-meter VISTA telescope and VIRCam \citep{Sutherland15} for near-infrared (NIR) photometry.~The current NGFS survey footprint covers 50~deg$^2$ with 19 DECam tiles of 2.65 deg$^2$ each, and detects point-sources at S/N~$=5$ down to $u'\!=\!26.5$, $g'\!=\!26.1$, $i'\!=\!25.3$, $J\!=\!24.0$ and $K_s\!=\!23.3$ AB mag, which corresponds to the GC luminosity function (GCLF) turnover at $M_V\!\simeq\!-7.4$ mag \citep[e.g.,][]{rejkuba12}.

Details on the reduction process and photometry will be provided in a forthcoming paper (Puzia et al.~2018,~{\it in prep.}).~In the following, we give a brief summary of the main survey characteristics.~The DECam data have been processed using the basic calibrated images delivered by the DECam Community pipeline \citep[v2.5.0][]{Valdes14}, which were corrected for bias, flat fielding, and image crosstalk.~In subsequently steps, we applied our custom background subtraction strategy.~The astrometry, photometric calibration and stacking were performed with the software SCAMP \citep[v2.2.6,][]{Bertin06}, Source Extractor \citep[SE, v2.19.5,][]{Bertin96} and SWARP \citep[v2.19.5][]{Bertin02}. The reference stars are from the 2MASS Point Source Catalog \citep{Skru06} and the SDSS stripe 82 standard star frames \citep{abaz09}.~Our optical-passband photometry was also cross-validated, using previous catalogs of the same sky area, specifically with the data provided in the work of \cite{Kim13}, who obtained photometry in the $U, B, V,$ and $I$ passbands, taken with Mosaic\,II camera at the 4-meter Blanco telescope at CTIO.~The average seeing on the stacked images is $2.06\pm0.09$\arcsec, $1.38\pm0.06$\arcsec, and $1.23\pm0.02$\arcsec\ in the $u'$, $g'$, and $i'$ filter, respectively.~The pixel scale of our final DECam image stacks is 0.263\arcsec\ which corresponds to 25.5\,pc at the distance of the Fornax cluster.

The near-IR VIRCAM observations were processed from scratch starting with the raw data, for which we developed a custom pipeline to do the basic calibration as well as the background modeling and subtraction, photometry, astrometric calibration, and the final image stacking.~This was done with the same software packages as described for the optical DECam data reduction.~The average seeing on the stacked $J$ and $K_s$ images is $0.87\pm0.03$\arcsec\ and $0.89\pm0.05$\arcsec, respectively.~The spatial resolution of the VISTA data is 0.34\arcsec/pix $=33$\,pc at the distance of Fornax.~The survey information and data reduction process of the near-infrared and the optical observations will be presented in a forthcoming paper (Puzia et al.~{\it in prep.}).

%%%%%%%%%%%%%%%%%%%%%%%%%%%%%%%%%%%%%%%%%%%%%%%%%%%%%%%%%%%%

\section{Analysis}
\subsection{Sample selection}
\label{txt:sample}
The nucleated dwarf galaxy sample of this work originates in the NGFS sample studied by \cite{eigenthaler2018} which consists of 258 dwarf galaxies in the inner 25\% of the virial radius of the Fornax cluster \citep[$R_{\rm vir}/4\simeq350$\,kpc;][]{drinkwater01}, centered on the cD galaxy NGC\,1399 (see Fig.~\ref{fig:spatial}).~Of this total dwarf galaxy population, we consider 75 ($29\%$) to be nucleated based on the following selection criteria: {\it (i)} the central object is located at the photometric centre of the spheroid or slightly offset ($\leq\!3\arcsec$), {\it (ii)} the nuclear cluster is detected in at least two filters, and {\it (iii)} the central object appears as a point source.~From the 75 nucleated dwarfs, we could further analyze 61 nuclei.~For six galaxies the surface brightness profile fits of their spheroid light did not converge to robust solutions, due to contamination by nearby objects or/and too low surface brightness values.~For eight of them that have good spheroid profile fits, we encountered unstable solutions in the very inner regions, mainly due to saturation of nearby sources, too-low S/N, and/or stacking artifacts in the area.~Since we are primarily interested in robust panchromatic photometry, we, therefore, exclude these objects from the subsequent analysis.~Two nucleated dwarf galaxies in our sample overlap with the ACS Fornax Cluster Survey sample \citep{Cote06}, FCC202 and FCC136, which are shown in Figure~\ref{fig:spatial} as red ellipses with green edges.

Even with sub-arcsecond seeing, the spatial resolution of DECam (1\,pix~$\!=\!25.5$\,pc) and VISTA (1\,pix~$\!=\!33$\,pc) all nuclei at the Fornax distance are unresolved point sources.~Nucleus detections in the central NGFS dataset for each filter reach these magnitudes:~$u'\!\simeq\!25.2$\,mag, $g'\!\simeq\!26.2$\,mag, $i'\!\simeq\!25$\,mag, $J\!\simeq\!21.9$\,mag, and $K_s\!\simeq\!23.1$\,mag.~The faintest nucleated dwarf galaxy detected in the Fornax central region has an absolute magnitude of $M_i\!\simeq\!-10.8$~mag.

\begin{figure}[t]
    %trim=left bottom right top
 \includegraphics[trim=4.8cm 1.35cm 4.4cm 1.5cm,clip,width=\columnwidth]{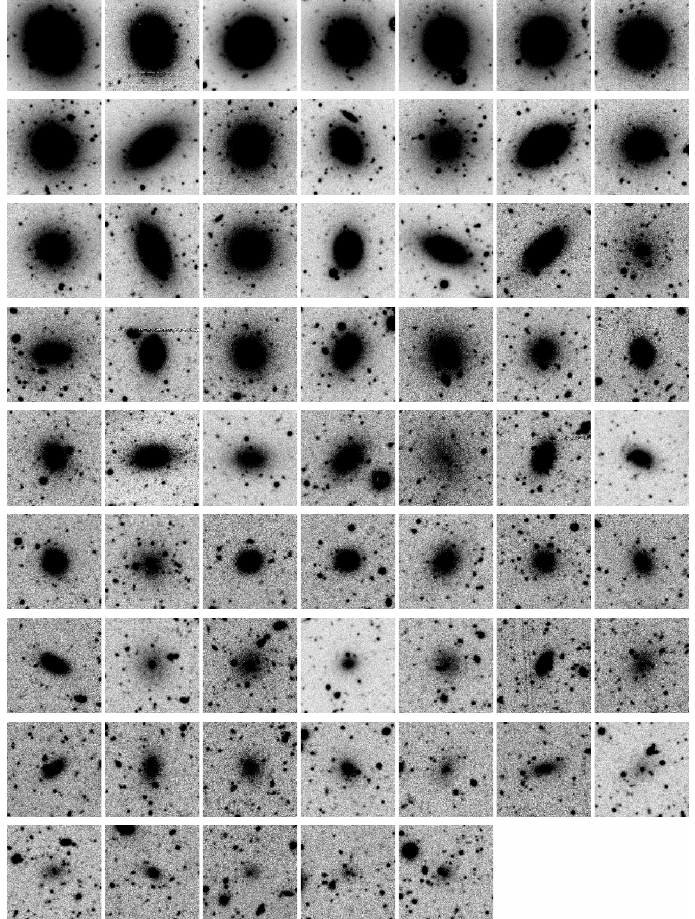}
 \caption{Postage stamp $g'$-band images for the 61 nucleated dwarfs, sorted by the $g'$-band luminosity of their spheroid component, which is well approximated by a S\'ersic profile \citep[see][]{eigenthaler2018}.~Each image has $200\!\times\!200\,{\rm pix}^2$, corresponding to $5.1\!\times\!5.1$\,kpc$^2$ at the Fornax distance.}
    \label{fig:sample}
\end{figure}

Figure~\ref{fig:sample} shows the gallery of $g'$-band images of our nucleated dwarf galaxy sample, ordered by decreasing luminosity of their spheroid component from top left to bottom right.~It is worth noting that the spheroid axis ratios and position angles of the dwarf galaxies are mostly consistent with rounded systems \citep[$\epsilon\!<\!0.55$; see][]{eigenthaler2018}.~Although the ellipticity distributions of the spheroid light components\footnote{This considers only the spheroid light component of the dwarf galaxy, excluding the nuclear star cluster.} of nucleated and non-nucleated dwarfs cover a similar range, \citeauthor{eigenthaler2018}~showed that the nucleated dwarfs are systematically round than their nucleus-devoid counterparts by $\Delta\epsilon\!\approx\!0.1$.~In addition, the spheroids of nucleated dwarfs have on average larger half-light radii ($\Delta r_{\rm eff}\approx0.2-0.3$\,kpc) than non-nucleated dwarfs.

We also observe that towards lower galaxy luminosities, the nucleus luminosity becomes more prominent compared to the luminosity of the galaxy spheroid, something that contrasts with the relation found for bright galaxies \citep{Turner12}.~Our main goal in this article is, therefore, to characterize the faint population of nuclear star clusters in the dwarf galaxies in the central Fornax region in terms of the information provided by their luminosities and colors using the broad SED coverage of our NGFS filter set.

\subsection{Spatial distribution}
Figure~\ref{fig:spatial} shows how our nucleated dwarf sample improves the sample size in terms of spatial coverage and luminosity range compared to previous studies in the same region, in particular those based on the ACSFCS observations which included nine nucleated dwarf galaxies.~The spatial distribution of the nucleated dwarfs in the central Fornax region follows the spatial distribution of the non-nucleated dwarf galaxy population, with a slight anisotropy in the east-west direction where the dwarf galaxy surface density appears to be mildly elevated compared to the north-south direction and with individual density peaks that follow the distribution of the giant galaxies \citep[see][]{Munoz15}.~Our earlier studies indicated that there may be a higher clustering of dwarf galaxies on length scales below $\sim\!100$\,kpc.~In galactocentric distance from NGC\,1399, the non-nucleated sample show a flat distribution out to $\sim\!350$\,kpc, meanwhile the nucleated dwarfs have a peak in surface number density at a cluster-centric radius of $\sim\!200$\,kpc and declining outwards.~Thus, data from the full NGFS survey footprint is required to understand the overall nucleated vs.~non-nucleated galaxy distribution in the Fornax galaxy cluster.~Some intriguing results on this topic from the sample presented in this work will be addressed in the discussion section below.

\subsection{Nucleation fraction}
\begin{figure}[t]
    %trim=left bottom right top
 \includegraphics[trim=0.2cm 0.25cm 0.2cm 0.15cm,clip,width=\columnwidth]{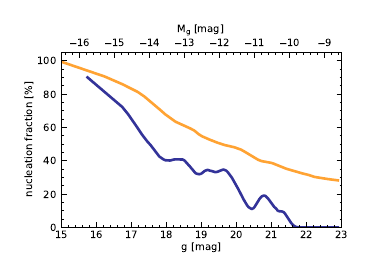}
 \caption{Nucleation fraction ($f_{\rm nuc}$) of Fornax dwarfs as a function of $g'$-band galaxy luminosity.~The orange line shows the cumulative distribution ($f_{\rm nuc}[<\!g']$), while the blue curve illustrates the differential relation ($\Delta f_{\rm nuc}/\Delta g'$).}
    \label{fig:fnuc}
\end{figure}

The nucleation fraction ($f_{\rm nuc}$) according to the previous work of \cite{Turner12} for the Fornax region is 72\% with galaxy luminosities of their sample reaching as low as $M_B\!\simeq\!-16$\,mag, thus, only covering the bright galaxy regime.~The NGFS nucleated dwarf galaxy sample extends this limit to $M_{g'}\!\simeq\!-10.5$\,mag.~In \cite{Munoz15}, we have found that $f_{\rm nuc}$ depends strongly on the galaxy luminosity.~In Figure~\ref{fig:fnuc} we use the NGFS dwarf galaxy sample to estimate cumulative and differential $f_{\rm nuc}$ as a function of galaxy luminosity in a window of 20 galaxies and smooth it using a Locally Weighted Scatterplot Smoothing (LOWESS) fit \citep[e.g.,][]{Clev81} to inspect the general trend.~The blue and the orange lines show the differential and cumulative $f_{\rm nuc}$ distributions, i.e.~$\Delta f_{\rm nuc}/\Delta g'$ and $f_{\rm nuc}(<\!g')$ respectively.~For the bright NGFS dwarf galaxies ($M_{g'}\!\la\!-16$\,mag) $f_{\rm nuc}$ reaches $\ga\!90\%$.~On the other hand, fainter galaxies show systematically lower $f_{\rm nuc}$ values, reaching zero at absolute magnitudes $M_{g'}\!\simeq\!-10$\,mag.~Although finding nucleated dwarfs towards fainter magnitudes depends on both the point-source detection limit (see values in Sect.~\ref{txt:sample}) and the surface brightness limit of our NGFS data \citep[$\langle\mu_{i'}\rangle_e\!\approx\!28$\,mag arcsec$^{-2}$, see][]{Munoz15, eigenthaler2018}, it is unclear whether ultra-low surface brightness spheroids are nucleated or not, as there may be detected nuclei in our NGFS point-source catalogues for which our photometry failed to detect the surrounding ultra-low surface brightness spheroid.~However, given the relatively faint point-source detection limit, we have a strong indication that for the sample of detected low surface brightness dwarf galaxy spheroids in Fornax, the nucleation stops at a well-defined galaxy luminosity ($M_{g'}\!\simeq\!-10$\,mag), corresponding to a galaxy stellar mass of $\log{\cal M}_\star\!\approx\!6.4$ \citep{eigenthaler2018}.

\subsection{Morphological decomposition of nucleus and spheroid}
\begin{figure}
    %trim=left bottom right top
\includegraphics[trim=2.3cm 1.9cm 2cm 0.5cm,clip,width=\columnwidth]{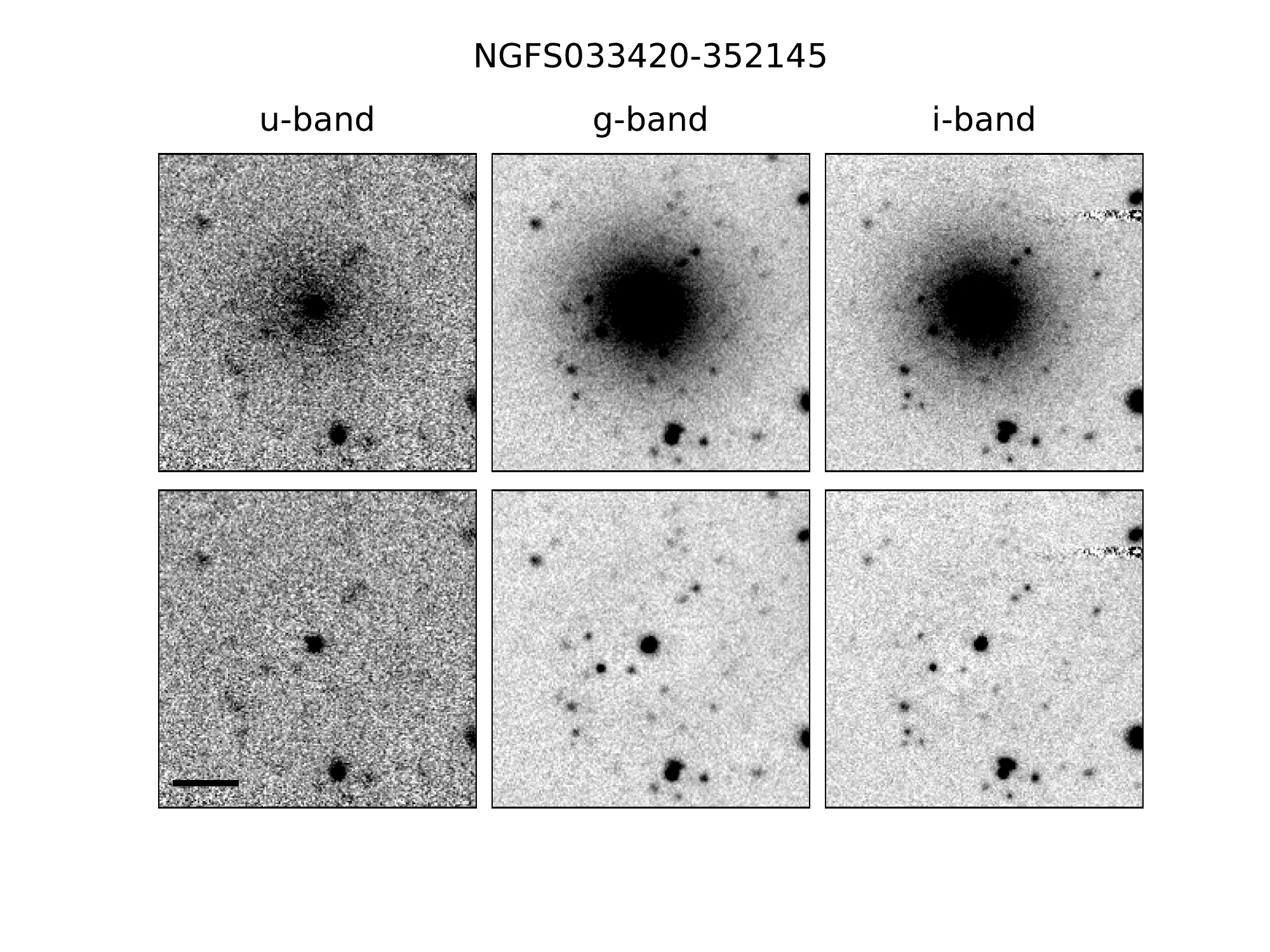}\\
\includegraphics[trim=2.3cm 2.2cm 2cm 0.5cm,clip,width=\columnwidth]{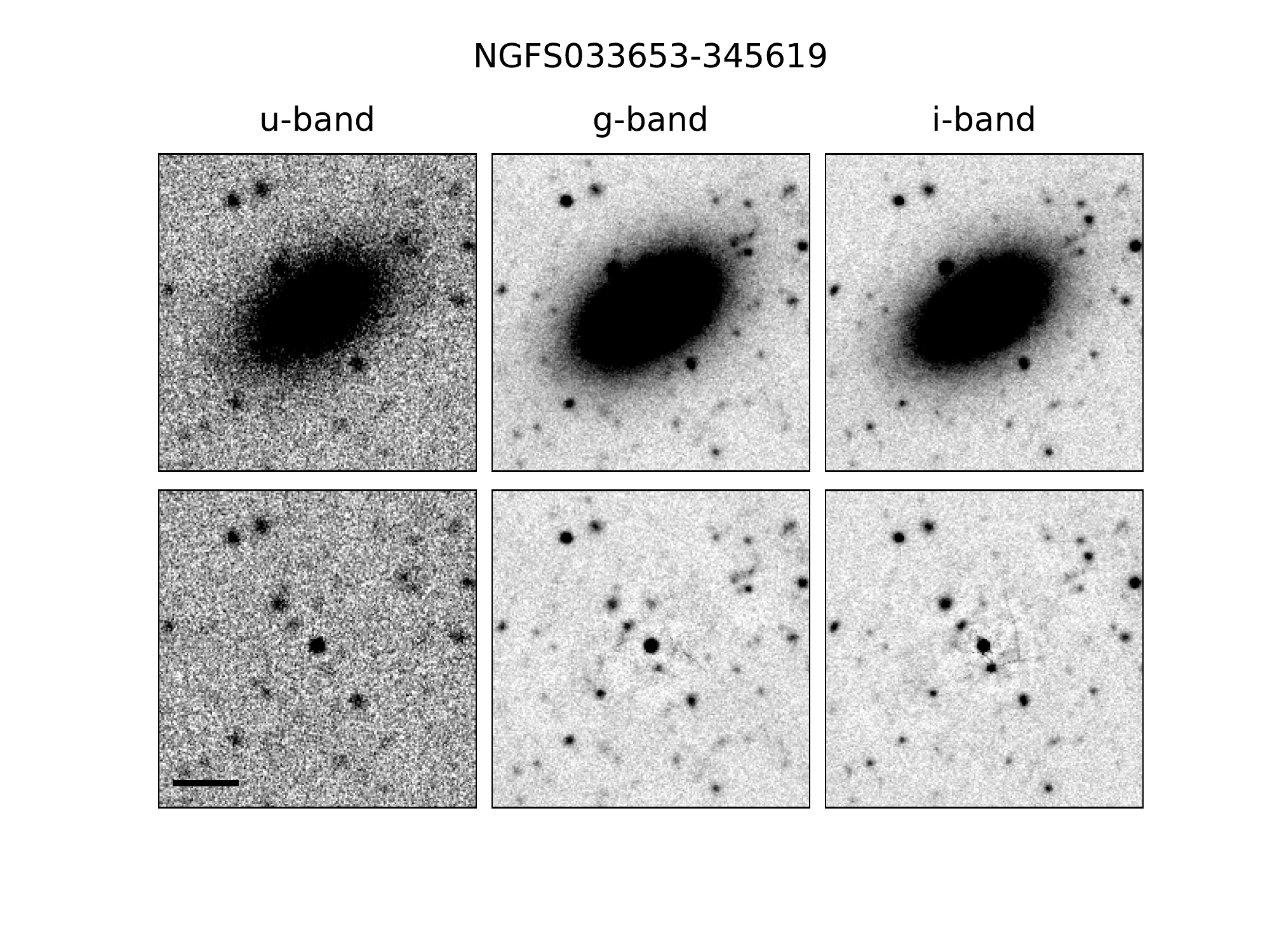}
   \caption{Ilustration of the galaxy fitting for two dwarf galaxies in the three optical bands, $u'g'i'$ from left to right, respectively.~The top rows show the original image and the bottom rows show the best residual images ({\sc galaxy -- model}), where the nuclear star cluster is deliberately left in the center, and is visible in all filters.~A scale bar (solid line) is shown at the bottom left image corresponding to $10.3\arcsec\,\hat{=}\,1$\,kpc.}
    \label{fig:ugi}
\end{figure}

Nuclei studies are affected by the host galaxy luminosity, and, therefore, it is necessary to subtract the galaxy spheroid light in order to study their intrinsic properties.~To accomplish this, we have developed an iterative approach to surface brightness profile fitting of dwarf galaxies using {\sc galfit} \citep[v3.0.5;][]{Peng02} and S\'ersic models in the $u'g'i'$ bands \citep[for more details, see][]{eigenthaler2018} and in the $JK_s$ bands.~The process to fit the light of a nucleated dwarf requires more iterations to achieve the best residual image, where the nucleus and spheroid are clearly separated.~The procedure to fit dwarf galaxies is as follows: {\it i)} a cutout image centered in the dwarf galaxy is created with size of $105\arcsec\!\times\!105\arcsec$ ($10.2$\,kpc~$\times10.2$\,kpc); {\it ii)} a mask image is created to cover all the sources around the dwarf, thus, creating a "dwarf-only" image; {\it iii)} {\sc galfit} is run over the cutout image using a PSF model (created with {\sc PSFex}) and the mask.~If the nucleus is present, a compact stellar object is left at or near the dwarf galaxy center; {\it iv)} the residual image is used to construct a new mask with the nucleus included; {\it v)} the steps from {\it ii)} to {\it iv)} are repeated at least three times to obtain the best nucleus-spheroid decomposition, leaving a residual image including the nucleus only.~Examples of the galaxy fitting process for two dwarf galaxies with different spheroid surface brightness levels and ellipticities are shown in Figure~\ref{fig:ugi}.~The dwarf galaxy images in the $u'g'i'$ filters are shown in the top-row panels, while the bottom-row panels show the corresponding residual images after the subtraction of the spheroid component.~The nucleus of each dwarf is clearly visible in the center of each residual image.~In several cases, other compact objects are found near the nucleus, which could be potential satellite globular clusters.~The analysis of the nucleus neighborhood and its constituent stellar populations will be presented in a forthcoming paper (Ordenes-Brice\~no et al., {\it in prep.}).

\subsection{Photometry}\label{phot_sect}
After the spheroid-nucleus fitting procedure we run SE with the corresponding PSF model and generate a catalog with the PSF photometry in all filters for each nucleus.~From the 61 dwarf galaxies (see Fig.~\ref{fig:sample}), 60 nuclei remain with reliable $i'$-band, 59 with $g'$-band, 28 with $u'$- and $K_s$-band photometry, 43 have good $J$-band.~The reasons for this inhomogeneity are partially or non-overlapping DECam and VISTA field of views, bad S/N and/or stacking in that area due to small overlap of the individual images, saturated or simply too faint nuclei.~Photometry in $u'g'i'JK_s$ is available for 28 nuclei.~Table~\ref{tab:nsc_parameters} shows the coordinates, photometry, and stellar masses for the 61 nuclei.~The photometry is corrected for foreground Galactic extinction with values taken from the latest \cite{Schf11} recalibration of the \cite{Schl98} dust reddening maps.~Reddening values for the different filters are calculated assuming the \cite{Fitz99} reddening law with $R_V\!=\!3.1$.~The average foreground extinction towards the central region of Fornax measured according to the brightest galaxies, i.e.~$A_{u}\!=\!0.056$, $A_{g}\!=\!0.043$, $A_{i}\!=\!0.022$, $A_{J}\!=\!0.009$, $A_{K_s}\!=\!0.004$.

\startlongtable
\begin{deluxetable*}{lcccccccc}
\tabletypesize{\scriptsize}
\tablecaption{Nuclear star clusters parameters\label{tab:nsc_parameters}}
\tablehead{
\colhead{Nucleus} & \colhead{RA (J2000)} & \colhead{Dec(J2000)} & \colhead{$u'$} & \colhead{$g'$} & \colhead{$i'$} & \colhead{$J$}  & \colhead{$K_s$} & \colhead{$\log({\cal M}_{*,{\rm nucleus}}$)}\\
\colhead{} & \colhead{\it hh:mm:ss.ss} & \colhead{\it dd:mm:ss.s} & \colhead{[AB mag]} & \colhead{[AB mag]} & \colhead{[AB mag]} & \colhead{[AB mag]}  & \colhead{[AB mag]} & \colhead{[$M_{\odot}$]}}
\startdata 
  NGFS033322-353620n & 03:33:22.19 &$-$35:36:20.2   &	     -           & 24.078$\pm$0.023  & 23.368$\pm$0.023  &          -	     &          -	   &  5.370$_{-0.008} ^{+0.009}$ \\
  NGFS033332-353942n & 03:33:32.16 &$-$35:39:42.3   &	     -           & 24.543$\pm$0.033  & 24.072$\pm$0.044  &          -	     &          -	   &  5.184$_{-0.013} ^{+0.013}$ \\
  NGFS033346-353455n & 03:33:46.05 &$-$35:34:56.0   &	     -           & 25.335$\pm$0.084  &        -          &          -	     &          -	   &             -               \\
  NGFS033412-351343n & 03:34:12.18 &$-$35:13:42.6   &	     -           & 25.046$\pm$0.052  & 23.735$\pm$0.036  &  23.222$\pm$0.101 &          -	   &  5.490$_{-0.041} ^{+0.137}$ \\
  NGFS033420-352145n & 03:34:20.17 &$-$35:21:44.7   &	     -           & 21.712$\pm$0.004  & 21.050$\pm$0.004  &  20.740$\pm$0.010 &          -	   &  6.310$_{-0.249} ^{+0.227}$ \\
  NGFS033444-355141n & 03:34:44.17 &$-$35:51:41.4   &	23.173$\pm$0.034 & 22.169$\pm$0.008  & 21.490$\pm$0.006  &  21.289$\pm$0.014 &  21.895$\pm$0.053   &  6.087$_{-0.269} ^{+0.284}$ \\
  NGFS033446-345334n & 03:34:46.13 &$-$34:53:34.2   &	     -           & 25.351$\pm$0.087  & 24.025$\pm$0.062  &          -	     &          -	   &  5.395$_{-0.066} ^{+0.059}$ \\
  NGFS033453-353411n & 03:34:52.74 &$-$35:34:10.7   &	     -           & 25.883$\pm$0.116  & 24.942$\pm$0.096  &          -	     &          -	   &  4.804$_{-0.087} ^{+0.106}$ \\
  NGFS033456-351127n & 03:34:56.49 &$-$35:11:27.0   &	24.360$\pm$0.085 & 23.195$\pm$0.015  & 22.494$\pm$0.014  &  22.415$\pm$0.041 &  22.520$\pm$0.096   &  5.658$_{-0.261} ^{+0.215}$ \\
  NGFS033458-350235n & 03:34:58.21 &$-$35:02:33.9   &	     -           & 22.424$\pm$0.007  & 21.736$\pm$0.007  &  21.533$\pm$0.024 &          -	   &  6.010$_{-0.253} ^{+0.240}$ \\
  NGFS033512-351923n & 03:35:11.50 &$-$35:19:22.6   &	25.020$\pm$0.162 & 23.638$\pm$0.018  & 22.997$\pm$0.018  &  22.775$\pm$0.052 &  22.724$\pm$0.101   &  5.460$_{-0.267} ^{+0.354}$ \\
  NGFS033524-362150n & 03:35:24.08 &$-$36:21:50.7   &	     -           &        -          & 19.645$\pm$0.004  &  19.555$\pm$0.008 &  19.632$\pm$0.013   &  6.562$_{-0.230} ^{+0.118}$ \\
  NGFS033525-361044n & 03:35:25.20 &$-$36:10:44.2   &	24.796$\pm$0.113 & 23.662$\pm$0.020  & 22.911$\pm$0.019  &  22.390$\pm$0.049 &  18.202$\pm$0.003   &  5.655$_{-0.281} ^{+0.229}$ \\
  NGFS033543-353051n & 03:35:42.79 &$-$35:30:50.7   &	24.258$\pm$0.066 & 23.245$\pm$0.019  & 22.587$\pm$0.013  &  22.362$\pm$0.033 &  22.849$\pm$0.108   &  5.636$_{-0.262} ^{+0.273}$ \\
  NGFS033546-355921n & 03:35:46.30 &$-$35:59:21.4   &	22.322$\pm$0.013 & 21.393$\pm$0.003  & 20.679$\pm$0.003  &  20.410$\pm$0.008 &  20.973$\pm$0.024   &  6.405$_{-0.268} ^{+0.290}$ \\
  NGFS033604-352320n & 03:36:04.05 &$-$35:23:19.7   &	23.968$\pm$0.060 & 22.917$\pm$0.011  & 22.191$\pm$0.013  &          -	     &  21.973$\pm$0.054   &  5.759$_{-0.252} ^{+0.370}$ \\
  NGFS033624-355442n & 03:36:23.64 &$-$35:54:40.8   &	24.483$\pm$0.086 & 23.405$\pm$0.013  & 22.664$\pm$0.012  &  22.421$\pm$0.048 &  23.081$\pm$0.164   &  5.627$_{-0.276} ^{+0.254}$ \\
  NGFS033628-351239n & 03:36:27.96 &$-$35:12:38.4   &	     -           & 24.569$\pm$0.044  & 23.847$\pm$0.042  &  23.251$\pm$0.090 &          -	   &  5.340$_{-0.297} ^{+0.433}$ \\
  NGFS033632-362537n & 03:36:32.21 &$-$36:25:37.3   &	     -           & 22.419$\pm$0.007  & 21.653$\pm$0.006  &  21.423$\pm$0.057 &          -	   &  6.087$_{-0.159} ^{+0.161}$ \\
  NGFS033637-352309n & 03:36:37.27 &$-$35:23:09.1   &	22.412$\pm$0.015 & 21.366$\pm$0.005  & 20.699$\pm$0.0032 &  20.320$\pm$0.006 &  20.909$\pm$0.020   &  6.438$_{-0.277} ^{+0.254}$ \\
  NGFS033653-345619n & 03:36:53.26 &$-$34:56:18.1   &	23.056$\pm$0.021 & 21.836$\pm$0.005  & 21.093$\pm$0.0038 &  20.806$\pm$0.012 &          -	   &  6.202$_{-0.255} ^{+0.346}$ \\
  NGFS033657-355011n & 03:36:57.12 &$-$35:50:11.4   &	     -           & 25.144$\pm$0.060  & 24.266$\pm$0.0491 &          -	     &          -	   &  5.035$_{-0.045} ^{+0.065}$ \\
  NGFS033700-350816n & 03:36:59.85 &$-$35:08:15.4   &	     -           & 25.062$\pm$0.051  & 24.563$\pm$0.0676 &  23.917$\pm$0.182 &          -	   &  5.050$_{-0.303} ^{+0.099}$ \\
  NGFS033700-352035n & 03:36:59.83 &$-$35:20:36.0   &	23.493$\pm$0.029 & 22.538$\pm$0.007  & 21.914$\pm$0.0065 &  21.519$\pm$0.017 &  21.984$\pm$0.053   &  5.918$_{-0.263} ^{+0.318}$ \\
  NGFS033703-354802n & 03:37:03.42 &$-$35:48:02.1   &	     -           & 24.784$\pm$0.041  & 23.696$\pm$0.0295 &  22.995$\pm$0.073 &          -	   &  5.461$_{-0.270} ^{+0.139}$ \\
  NGFS033708-344353n & 03:37:08.16 &$-$34:43:52.4   &	24.208$\pm$0.067 & 22.791$\pm$0.010  & 22.121$\pm$0.0117 &  21.655$\pm$0.038 &  21.697$\pm$0.091   &  5.857$_{-0.274} ^{+0.273}$ \\
  NGFS033710-352312n & 03:37:10.04 &$-$35:23:11.8   &	     -           & 24.537$\pm$0.090  & 23.664$\pm$0.0386 &  22.914$\pm$0.059 &          -	   &  5.539$_{-0.282} ^{+0.233}$ \\
  NGFS033710-355317n & 03:37:10.35 &$-$35:53:16.9   &	     -           & 25.336$\pm$0.072  & 24.608$\pm$0.0793 &          -	     &          -	   &  4.867$_{-0.028} ^{+0.031}$ \\
  NGFS033718-354157n & 03:37:17.93 &$-$35:41:57.3   &	23.708$\pm$0.049 & 22.711$\pm$0.010  & 21.929$\pm$0.0082 &  21.439$\pm$0.018 &  21.835$\pm$0.047   &  5.972$_{-0.290} ^{+0.336}$ \\
  NGFS033727-355747n & 03:37:27.49 &$-$35:57:46.8   &	     -           & 24.945$\pm$0.046  & 25.017$\pm$0.0928 &          -	     &          -	   &  5.023$_{-0.020} ^{+0.018}$ \\
  NGFS033734-354945n & 03:37:34.04 &$-$35:49:44.9   &	24.632$\pm$0.095 & 23.451$\pm$0.015  & 22.754$\pm$0.0142 &  22.449$\pm$0.048 &  22.879$\pm$0.133   &  5.565$_{-0.263} ^{+0.315}$ \\
  NGFS033742-343816n & 03:37:41.97 &$-$34:38:15.7   &	     -           & 21.601$\pm$0.004  & 20.961$\pm$0.0031 &  20.670$\pm$0.028 &          -	   &  6.203$_{-0.204} ^{+0.120}$ \\
  NGFS033817-353028n & 03:38:16.64 &$-$35:30:27.5   &	24.332$\pm$0.089 & 23.276$\pm$0.017  & 22.637$\pm$0.0145 &  22.458$\pm$0.043 &  22.588$\pm$0.094   &  5.586$_{-0.260} ^{+0.274}$ \\
  NGFS033837-355002n & 03:38:36.63 &$-$35:50:02.1   &	     -           & 25.875$\pm$0.105  & 24.899$\pm$0.0889 &  22.174$\pm$0.034 &          -	   &  5.671$_{-0.270} ^{+0.349}$ \\
  NGFS033837-355502n & 03:38:37.23 &$-$35:55:01.5   &	     -           & 26.191$\pm$0.153  & 25.057$\pm$0.1063 &  22.630$\pm$0.057 &          -	   &  5.440$_{-0.272} ^{+0.370}$ \\
  NGFS033838-354527n & 03:38:37.66 &$-$35:45:27.6   &	23.502$\pm$0.031 & 22.544$\pm$0.007  & 21.866$\pm$0.0057 &  21.346$\pm$0.015 &  22.084$\pm$0.058   &  5.997$_{-0.287} ^{+0.208}$ \\
  NGFS033845-351600n & 03:38:45.40 &$-$35:15:59.6   &	24.432$\pm$0.074 & 23.578$\pm$0.015  & 22.883$\pm$0.0141 &  22.341$\pm$0.038 &  23.053$\pm$0.156   &  5.611$_{-0.295} ^{+0.176}$ \\
  NGFS033854-344932n & 03:38:54.26 &$-$34:49:32.4   &	     -           & 24.096$\pm$0.021  & 23.307$\pm$0.0243 &          -	     &          -	   &  5.377$_{-0.009} ^{+0.016}$ \\
  NGFS033906-360557n & 03:39:05.77 &$-$36:05:56.2   &	22.584$\pm$0.018 & 21.526$\pm$0.004  & 20.786$\pm$0.003  &  20.440$\pm$0.008 &  21.080$\pm$0.030   &  6.368$_{-0.270} ^{+0.278}$ \\
  NGFS033913-352217n & 03:39:13.32 &$-$35:22:16.8   &	22.080$\pm$0.014 & 21.274$\pm$0.005  & 20.526$\pm$0.0037 &  20.229$\pm$0.006 &  20.775$\pm$0.018   &  6.486$_{-0.275} ^{+0.269}$ \\
  NGFS033920-354329n & 03:39:19.69 &$-$35:43:28.6   &	22.946$\pm$0.024 & 22.073$\pm$0.006  & 21.356$\pm$0.0055 &  21.036$\pm$0.011 &  21.840$\pm$0.045   &  6.156$_{-0.275} ^{+0.266}$ \\
  NGFS033950-353122n & 03:39:50.08 &$-$35:31:22.1   &	24.062$\pm$0.074 & 23.256$\pm$0.014  & 22.706$\pm$0.015  &  22.249$\pm$0.032 &  22.985$\pm$0.126   &  5.640$_{-0.307} ^{+0.289}$ \\
  NGFS033955-353943n & 03:39:55.44 &$-$35:39:42.9   &	     -           & 25.251$\pm$0.063  & 24.227$\pm$0.0465 &          -	     &          -	   &  5.127$_{-0.049} ^{+0.064}$ \\
  NGFS033956-353721n & 03:39:56.45 &$-$35:37:20.7   &	24.124$\pm$0.055 & 23.166$\pm$0.011  & 22.419$\pm$0.01   &  22.106$\pm$0.033 &  22.604$\pm$0.090   &  5.699$_{-0.266} ^{+0.299}$ \\
  NGFS034001-344323n & 03:40:00.56 &$-$34:43:23.3   &	23.817$\pm$0.057 & 23.103$\pm$0.013  & 22.244$\pm$0.011  &  22.302$\pm$0.067 &  22.264$\pm$0.169   &  5.798$_{-0.259} ^{+0.214}$ \\
  NGFS034010-355011n & 03:40:09.77 &$-$35:50:10.1   &	     -           & 24.234$\pm$0.029  & 23.735$\pm$0.03   &  23.271$\pm$0.128 &          -	   &  5.214$_{-0.247} ^{+0.226}$ \\
  NGFS034019-361850n & 03:40:19.35 &$-$36:18:49.9   &	     -           & 24.579$\pm$0.033  & 23.819$\pm$0.0315 &          -	     &          -	   &  5.127$_{-0.049} ^{+0.064}$ \\
  NGFS034023-351636n & 03:40:23.52 &$-$35:16:35.7   &	22.556$\pm$0.016 & 21.593$\pm$0.004  & 20.832$\pm$0.0028 &  20.676$\pm$0.008 &  21.187$\pm$0.029   &  6.380$_{-0.284} ^{+0.215}$ \\
  NGFS034027-362957n & 03:40:26.99 &$-$36:29:55.8   &	     -           & 21.709$\pm$0.006  & 20.978$\pm$0.0048 &          -	     &          -	   &  6.318$_{-0.002} ^{+0.002}$ \\
  NGFS034031-355241n & 03:40:30.65 &$-$35:52:40.7   &	     -           &        -          & 23.878$\pm$0.0505 &  22.741$\pm$0.068 &          -	   &  5.735$_{-0.272} ^{+0.091}$ \\
  NGFS034034-350122n & 03:40:33.83 &$-$35:01:22.3   &	22.702$\pm$0.038 & 21.597$\pm$0.006  & 20.768$\pm$0.0041 &  20.656$\pm$0.011 &  21.121$\pm$0.036   &  6.421$_{-0.271} ^{+0.181}$ \\
  NGFS034038-360716n & 03:40:37.76 &$-$36:07:16.4   &	     -           & 24.655$\pm$0.038  & 23.903$\pm$0.0386 &          -	     &          -	   &  5.127$_{-0.049} ^{+0.064}$ \\
  NGFS034044-361108n & 03:40:43.85 &$-$36:11:07.7   &	     -           & 24.081$\pm$0.025  & 23.509$\pm$0.0293 &  23.049$\pm$0.191 &          -	   &  5.235$_{-0.222} ^{+0.153}$ \\
  NGFS034050-354454n & 03:40:50.40 &$-$35:44:54.4   &	20.788$\pm$0.005 & 19.603$\pm$0.002  & 18.718$\pm$0.0011 &  18.389$\pm$0.002 &  18.771$\pm$0.003   &  7.255$_{-0.261} ^{+0.289}$ \\
  NGFS034101-354434n & 03:41:00.78 &$-$35:44:33.2   &	21.995$\pm$0.010 & 20.959$\pm$0.003  & 20.272$\pm$0.0019 &  19.993$\pm$0.007 &  20.546$\pm$0.018   &  6.558$_{-0.264} ^{+0.305}$ \\
  NGFS034107-353052n & 03:41:07.23 &$-$35:30:51.9   &	     -           & 24.355$\pm$0.035  & 23.706$\pm$0.0335 &  23.204$\pm$0.115 &          -	   &  5.310$_{-0.289} ^{+0.143}$ \\
  NGFS034113-350932n & 03:41:12.87 &$-$35:09:31.3   &	21.879$\pm$0.010 & 20.890$\pm$0.003  & 20.223$\pm$0.0024 &  20.212$\pm$0.010 &  20.674$\pm$0.029   &  6.490$_{-0.015} ^{+0.081}$ \\
  NGFS034135-352625n & 03:41:35.06 &$-$35:26:24.3   &	     -           & 23.736$\pm$0.018  & 22.882$\pm$0.0143 &          -	     &          -	   &  6.318$_{-0.002} ^{+0.002}$ \\
  NGFS034217-353227n & 03:42:17.25 &$-$35:32:26.6   &	     -           & 20.562$\pm$0.003  & 19.801$\pm$0.0018 &          -	     &          -	   &  6.780$_{-0.001} ^{+0.001}$ \\
  NGFS034218-352819n & 03:42:17.83 &$-$35:28:18.7   &	     -           & 25.447$\pm$0.082  & 24.545$\pm$0.0717 &          -	     &          -	   &  4.938$_{-0.069} ^{+0.076}$ \\
  NGFS034225-353541n & 03:42:25.22 &$-$35:35:41.5   &	     -           & 24.208$\pm$0.024  & 23.421$\pm$0.0212 &          -	     &          -	   &  5.127$_{-0.049} ^{+0.064}$ \\
\enddata
%\tablecomments{$^{a}$: coordinates are from the spheroid.}
\end{deluxetable*}

\normalsize

%%%%%%%%%%%%%%%%%%%%%%%%%%%%%%%%%%%%%%%%%%%%%%%%%%%%%%%%%%%%

\section{Results}

\subsection{Color-magnitude and color-color diagrams}
\begin{figure}[t!]
     \includegraphics[trim=0.cm 0.6cm 0.cm 0.8cm,clip,width=1.02\columnwidth]{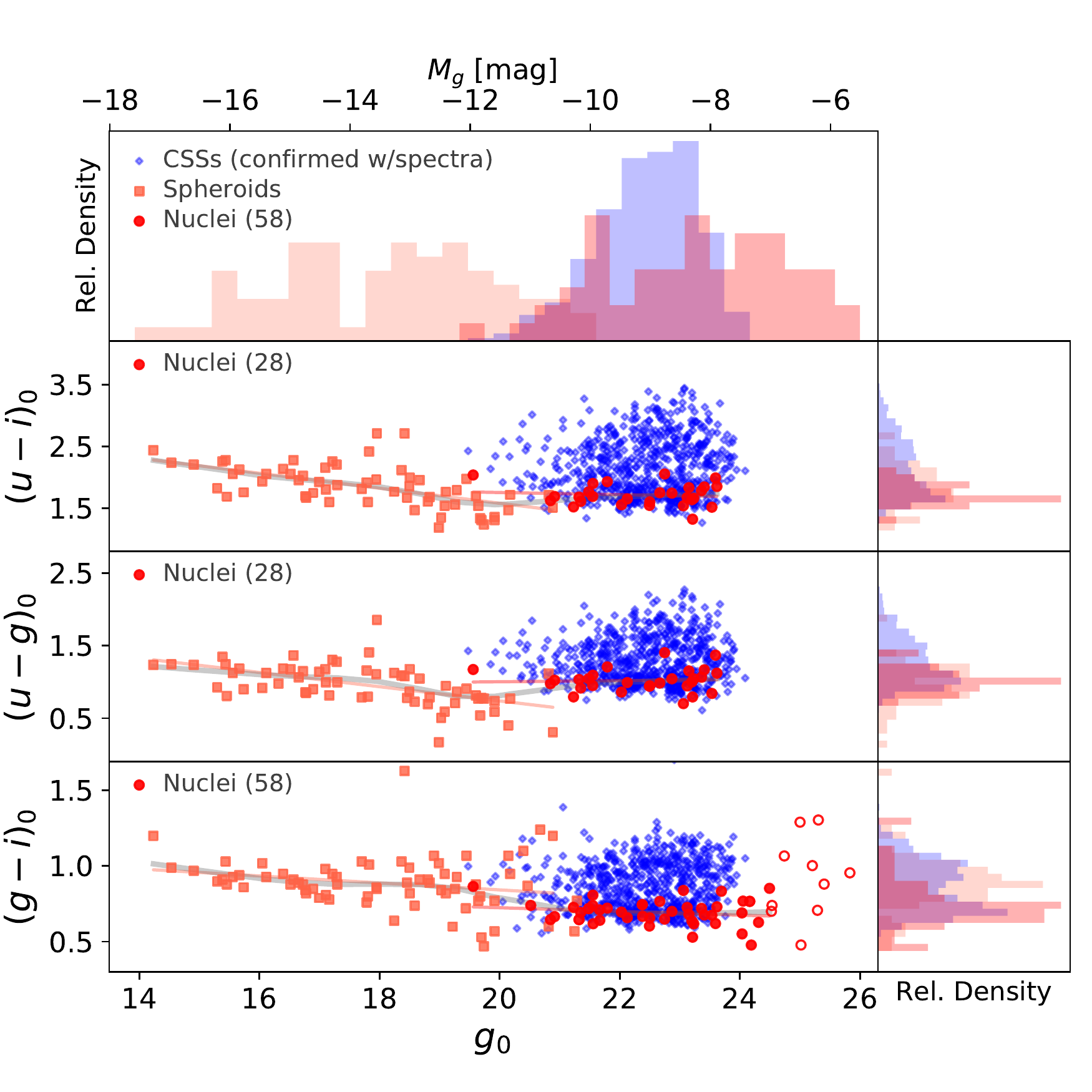}
     \caption{Color-magnitude diagrams for the NGFS sample nuclei shown as red circles.~Blue diamonds are CSSs that were confirmed by radial velocities, taken from \cite{Witt16}, and orange squares mark the spheroid colors and luminosities of the host dwarf galaxies in which the nuclei of our sample were found.~Linear relations show the weighted least-square fits to the nuclei and spheroid color-magnitude relations (see text for details), while the grey curve is a LOWESS fit to the combined nuclei+spheroid sample.}
   \label{fig:cmd}
\end{figure}

Figure~\ref{fig:cmd} illustrates the color-magnitude diagrams (CMDs) in the filter combinations $(u'\!-\!i')_0$, $(u'\!-\!g')_0$ and $(g'\!-\!i')$ vs.~$g'_0$.~The nuclei of this work are shown as red circles and their host galaxy spheroid components as orange squares.~For comparison, we plot also radial-velocity confirmed compact stellar systems (CSSs, i.e.~GCs and UCDs) near the cD galaxy, NGC\,1399 from the clean compilation catalog in \cite{Witt16} as blue symbols.

The nuclei occupy the bluest parts of the CSS distribution in all three colors with mean values $\langle(u'-i')_{0,{\rm nuc}}\rangle\!=\!1.71\pm0.03$, $\langle(u'-g')_{0,{\rm nuc}}\rangle\!=\!1.02\pm0.03$, $\langle(g'-i')_{0,{\rm nuc}}\rangle\!=\!0.73\pm0.03$ compared to CSSs that cover a significantly broader range of colors with mean values $\langle(u'-i')_{0,{\rm CSSs}}\rangle\!=\!2.15\pm0.02$, $\langle(u'-g')_{0,{\rm CSSs}}\rangle\!=\!1.28\pm0.01$, $\langle(g'-i')_{0,{\rm CSSs}}\rangle\!=\!0.87\pm0.01$.~The broader distribution for CSSs and their extension to redder colors is mainly due to their larger spread in metal content reaching super-solar values \citep[e.g.][]{kp98}.~On the other hand, both samples have similar luminosity distributions.~One of the key findings here is that the NGFS nuclei show a flat color-magnitude relation (CMR).~Whether this is due to no significant changes in the stellar population content as a function of stellar mass will be discussed below.~We point out that this is opposite to other results from studies focused on nuclei in the brighter nucleated dwarf galaxy regime, where a CMR for nuclei was found in the magnitude range $-16\!\ga\!M_B\!\ga\!-18$\,mag, while for nucleated galaxies brighter than $M_B\!\simeq\!-18$\,mag the relation becomes flatter again  \citep{Cote06, Turner12, Spengler17}.

\begin{deluxetable}{lccc}[!t]
\tabletypesize{\scriptsize}
\tablecaption{Nucleus and spheroid color-magnitude relations\label{tab:cmr}}
\tablehead{
\colhead{Linear wLSQ fits} & \colhead{$r$} & \colhead{$p$} & \colhead{$\sigma$} }
\startdata
$(u'\!-\!i')_{\rm nuc}=-0.017\,g'+2.094$  & $-0.103$ & $0.602$ & $0.032$ \\
$(u'\!-\!i')_{\rm sph}=-0.121\,g'+4.008$  & $-0.582$ & $8.511\cdot10^{-7}$ & $0.022$ \\
$(u'\!-\!g')_{\rm nuc}=+0.008\,g'+0.848$  & $+0.050$ & $0.800$ & $0.030$ \\
$(u'\!-\!g')_{\rm sph}=-0.098\,g'+2.704$  & $-0.560$ & $2.744\cdot10^{-6}$ & $0.019$ \\
$(g'\!-\!i')_{\rm nuc}=-0.012\,g'+0.969$  & $-0.175$ & $0.267$ & $0.011$ \\
$(g'\!-\!i')_{\rm sph}=-0.023\,g'+1.303$  & $-0.229$ & $0.076$  & $0.013$ \vspace{1mm} \\
\enddata
\tablecomments{The left column shows the weighted least-square fits for nuclei and spheroids in various filter combinations (see Fig.~\ref{fig:cmd}), the second column gives the correlation coefficient ($r$) and the next the $p$-value for the hypothesis that the CMR has zero slope, while the last column is the standard error of the gradient.~The relations for the spheroids are valid in the range $-18\!\la\! g'\!\la\!-10.5$, while the corresponding relations for the nuclei are valid in the range $-11\!\la\! g'\!\la\!-7.5$ mag.}
\end{deluxetable}

In comparison to the nuclei, the spheroid components of their host galaxies show a shallow but measurable CMR with the spheroid colors becoming redder for brighter systems.~Table~\ref{tab:cmr} summarizes the numerical properties of the weighted linear least-square fits to the CMR of nuclei and spheroids in various filter combinations.~The CMR of the red-sequence galaxies (including dwarfs) in Virgo and Fornax was recently shown to become flatter going from brightest galaxies towards the faint dwarf luminosity regime \citep{Roediger17, eigenthaler2018}.~The CMR is usually interpreted as a mass-metallicity relation (MZR) due to deeper potential wells retaining more metals produced by stars during the secular evolution of the galaxy compared to their less massive counterparts \citep[e.g.][and references therein]{Kod97, Trem04, Kewley08, Torrey17}.~In relation to the flat CMR of the nuclei we find at the overlap luminosity of $M_{g'}\!\simeq\!-11.0$\,mag an offset of $\delta(u'-i')_0\approx0.2$ and $\delta(u'-g')_0\approx0.31$, but a relatively small offset in the optical color of $\delta(g'-i')_0\approx0.13$.~These color offsets have implications for the differences in stellar population contents between nuclei and the surrounding galaxy spheroids and provide constraints for the formation mechanisms of galaxy nuclei and the  build-up of galaxies and CSSs in galaxy clusters.~We will come back to this point in the discussion section.

\begin{figure}[!b]
    %trim=left bottom right top
    \includegraphics[trim=0.1cm 1.5cm 1.7cm 2.95cm,clip,width=\columnwidth]{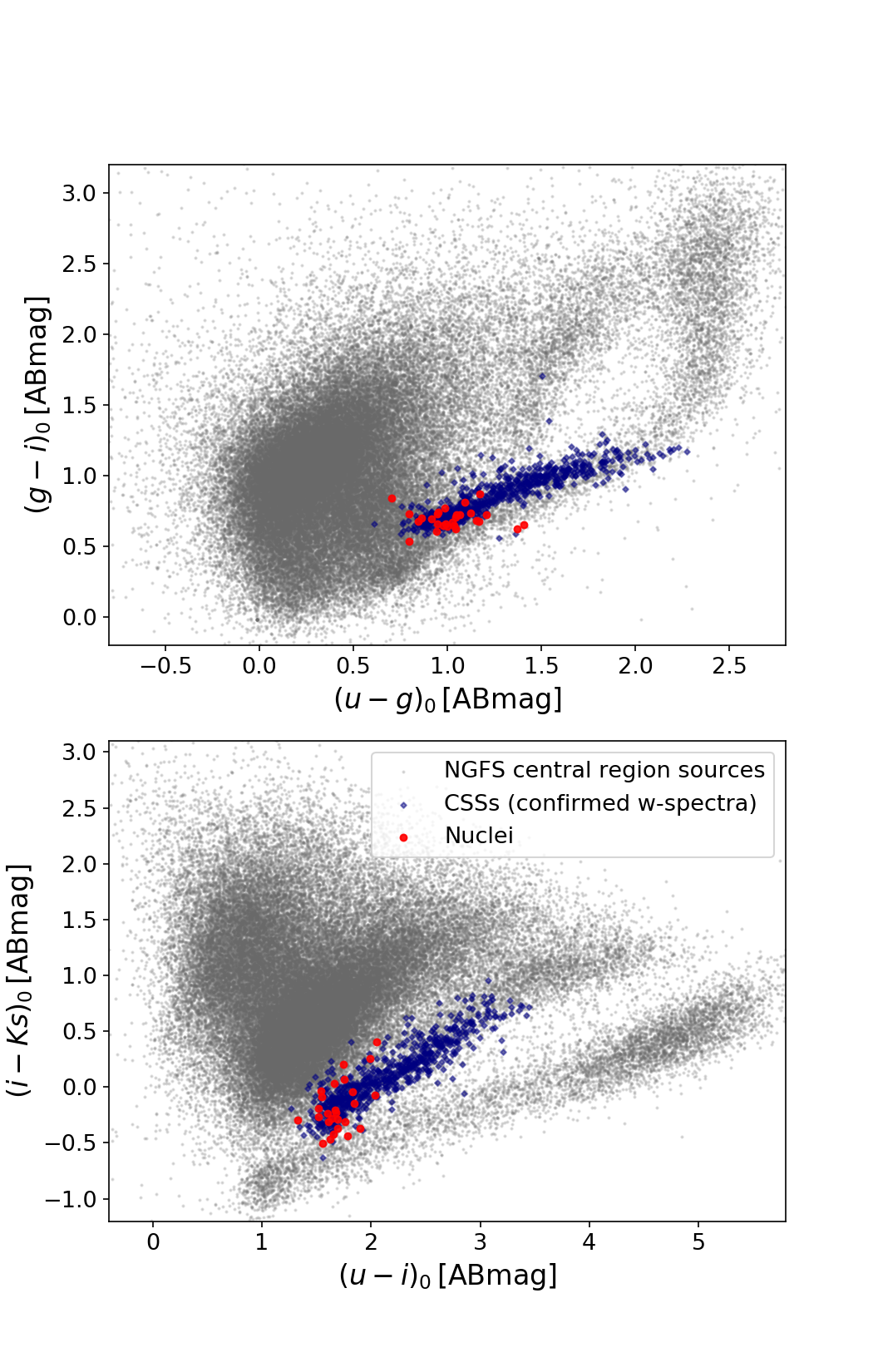}
     \caption{Color-color diagrams, $(u'\!-\!g')_0$ vs.~$(g'\!-\!i')_0$ (top panel) and $(u'\!-\!i')_0$ vs.~$(i'\!-\!K_s)_0$ (bottom panel), showing as gray dots all the NGFS central sources in the $R < R_{\rm vir}/4$ with PSF photometry.~The red filled circles represent the nuclei of this work and the blue diamonds show the confirmed CSSs in the surroundings of the cD galaxy NGC\,1399.~Compact stellar systems such as GCs, UCDs and nuclei lie in a very narrow region in color-color space, which makes the $u'g'i'$ and $u'i'K_s$ diagrams very powerful tools to select CSSs candidates.}
     \label{fig:ccd}
\end{figure}

Figure~\ref{fig:ccd} shows two color-color diagrams, i.e.,~$(u'\!-\!g')_0$ vs.~$(g'\!-\!i')_0$ or $u'g'i'$ (top panel) and $(u'\!-\!i')_0$ vs $(i'\!-\!K_s)_0$ or $u'i'K_s$ (bottom panel), where gray dots represent all the detected NGFS sources for which PSF-based photometry could be obtained in the central region of Fornax (Fig.~\ref{fig:spatial}).~These color-color diagrams, in particular the $u'i'K_s$ plane with its broad SED coverage, are powerful tools to distinguish among different object types, such as foreground stars, background galaxies and CSSs \citep{Munoz14}.~The upper-left cloud of objects in both diagrams shows the location of red-shifted background galaxies, while in the lower parts of the $u'g'i'$ diagram, a tight sequence holds for individual foreground stars and star clusters, but in the $u'i'K_s$ diagram even these objects are separated into two sequences.~Here the central sequence marks the star cluster sequence, as it is shown by the confirmed CSSs (blue symbols) from the very central parts of the Fornax cluster.~Our nuclei sample (red filled circles) is located in the same color-color region and confined to the bluest parts of the star cluster sequence, as was already seen in the color-magnitude diagrams (see Fig.~\ref{fig:cmd}).~The analysis of the complete star-cluster photometry catalog in the central Fornax region with the new GCs and UCDs candidates will be reported in a future work.~In the subsequent analysis we focus on the dwarf galaxy nuclei.

\subsection{Stellar mass estimates}
\label{txt:stellar_mass}
Stellar masses for our NGFS nuclei are estimated using a $\chi^2$ minimization approach to fit stellar population synthesis models to the photometric information from the NGFS filters $g'i'JK_s$, $g'i'J'$, or $g'i'$, according to the photometry available for each nucleus.~We exclude the $u'$-band photometry due to its sensitivity to very young stellar populations with low mass fractions and/or potential AGN emission components, if any.~The mass errors are estimated with Monte-Carlo simulations by drawing one thousand random values from a normal probability distribution function with a mean corresponding to the observed magnitude and a standard deviation equal to the magnitude error, and propagating these values through the following calculations.~SSP models from \citet[][hereafter BC03]{BC03} with the 2016 update\footnote{http://www.bruzual.org/$\sim$gbruzual/bc03/Updated\_version\_2016/}, MILES atlas \citep{sanc06}, and an initial mass function (IMF) from \cite{Kroupa01} are used to estimate luminosity weighted stellar masses.~We consider metallicities in the range $0.0001\leq Z/Z_{\odot}\leq 0.5$ and ages older than 1\,Gyr to avoid the stochasticity typically found at younger ages \citep[e.g.][]{cer02, cer04, fos10}.~The stellar mass distribution of our nuclei covers the range $\log({\cal M}_\star/M_\odot)=\!4.8\!-\!7.3$ with uncertainties ranging from $\sim$8\% to 43\% and a mean uncertainty of $\sim\!19\%$, propagated from the photometric errors.~We point out that there are systematic uncertainties of the mass estimates related to the choice of population synthesis models and the set of filters used to compute the mass-to-light conversion \citep[see][]{powalka16b, powalka17, zhang17}, which for our sample we estimate to be at most $\sim\!0.2$ dex.

\begin{figure}
%trim=left bottom right top
    \centering
    \includegraphics[trim=0.7cm 0.1cm 0.9cm 0.6cm,clip,width=\linewidth]{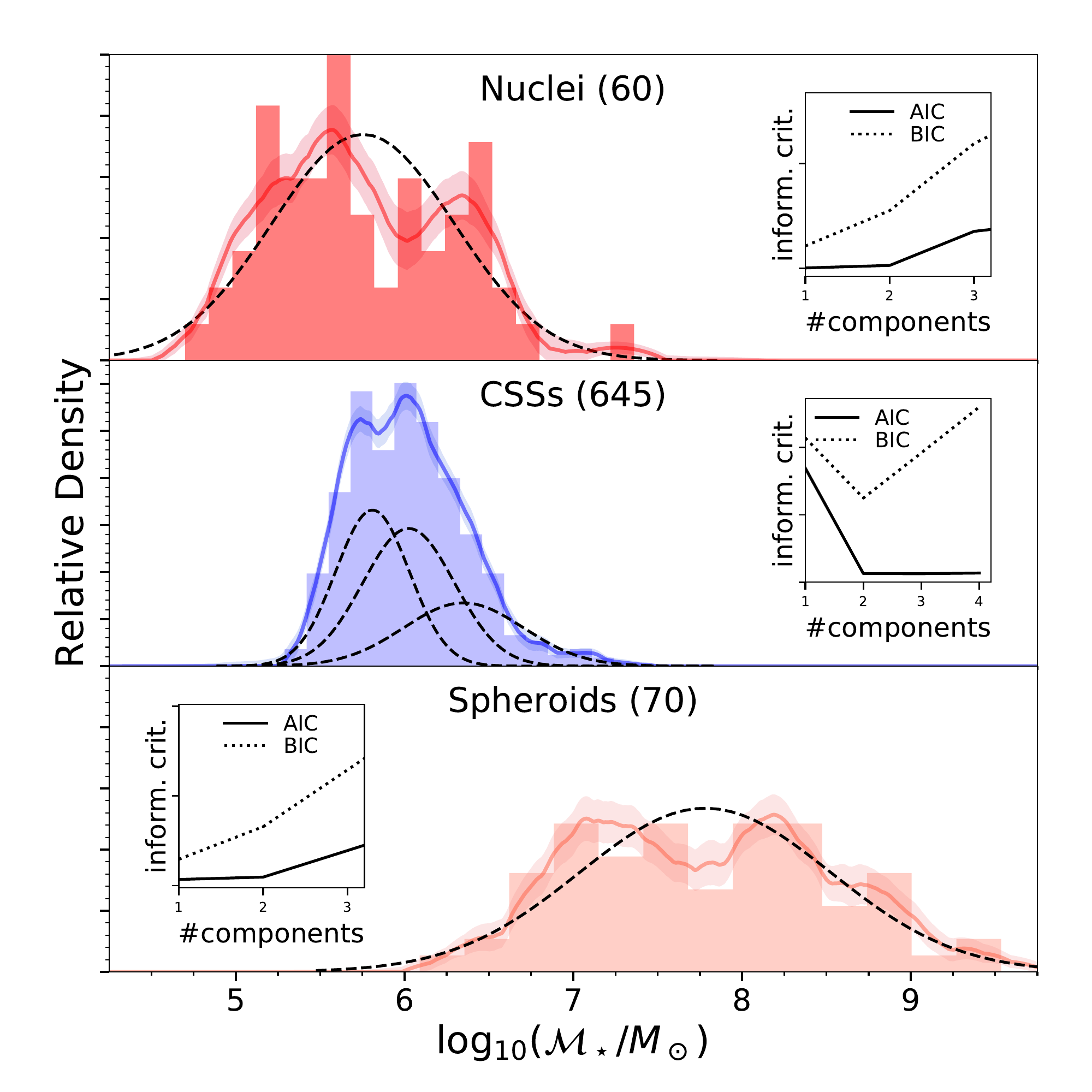}
     \caption{Mass distribution for the NGFS nuclei (top panel), the radial-velocity confirmed CSSs in Fornax (middle panel), and the dwarf spheroids (bottom panel).~The probability density distribution for each sample is overplotted using an Epanechnikov-kernel density estimate together with the 1-$\sigma$ uncertainty ranges.~The size of each sample is given in parentheses in the corresponding panel.~The inset figures show the Aikake and Schwarz's (Bayesian) information criteria \citep[AIC and BIC, see][]{astroML14}, which define the most likely number of Gaussian components for each distribution; here we use the AIC.~The corresponding Gaussians are indicated by the dashed curves.}
    \label{fig:massdistr}
\end{figure}

The stellar mass distribution for the NGFS nuclei is illustrated in Figure~\ref{fig:massdistr}, together with the distribution of the confirmed CSSs in the Fornax region, for which we estimate their masses with the same procedure applied to our nuclei, as well as the nucleated NGFS dwarf spheroids \footnote{Note that we consider the spheroid mass alone, i.e.~the nucleated spheroid mass does not include the nucleus.}, for which we use the mass measurements estimated in \cite{eigenthaler2018}, which are based on the parametrizations of the mass-to-light ratios as a function of various colors given in \cite{Bell03}.~We point out that our sample dwarf spheroids do not show any signs of star formation activity and, hence, their optical colors serve as good indicators for their stellar masses \citep[see][for the numerical accuracy of these conversions]{zhang17}.~An Epanechnikov-kernel probability density estimate (KDE) is overplotted for each distribution together with its 1-$\sigma$ uncertainties.

\begin{figure*}[ht!]
    %trim=left bottom right top
    \centering
    \includegraphics[trim=0cm 1.7cm 0.cm 3.2cm,clip,width=0.8\textwidth]{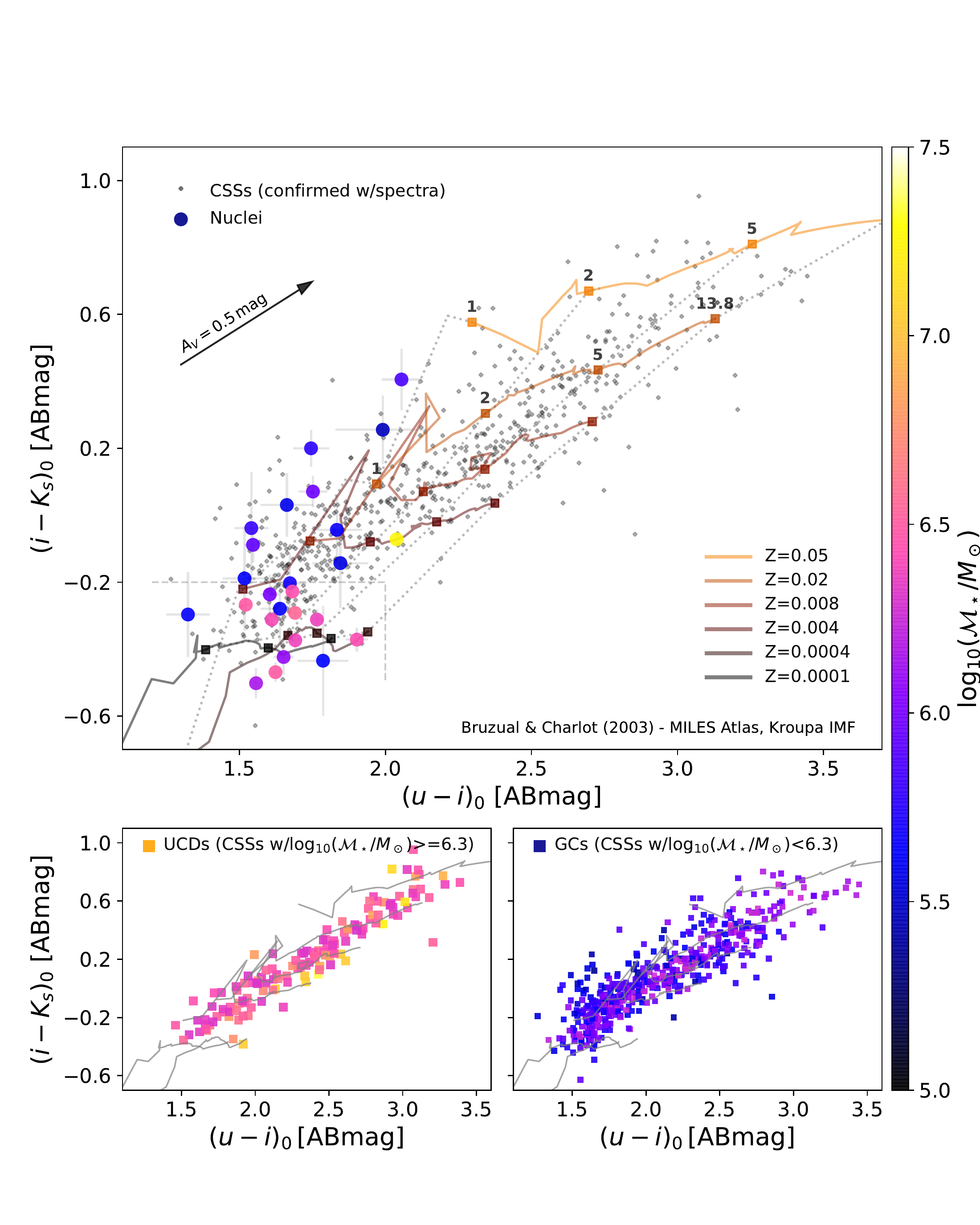}
    \caption{{\it Top panel:}$(u'\!-\!i)_0$ vs.~$(i'\!-\!K_s)_0$ color-color diagram with SSP models from BC03 using the MILES stellar library and a Kroupa IMF.~Iso-metallicity tracks are shown as solid lines, ranging from $Z\!=\!0.0001$ to $0.05$ (see legend).~Squares mark ages of $1, 2, 5,$ and $13.8$\,Gyr on each iso-metallicity track (see labels on the $Z\!=\!0.05$ and $Z_\odot$ track).~These points are connected by dotted iso-age lines.~Nuclei are color coded using their stellar masses as indicated by the colorbar.~The arrow in the top-left corner shows a reddening vector of $A_V\!=\!0.5$\,mag.~{\it Bottom panels:}~The left and the right panels indicate the color-color relations of UCDs and GCs, respectively.~Their stellar masses are color-coded on the same scale as for the top panel.~GC stellar masses range from $10^{5.4}$ to $10^{6.3}\,M_\odot$.~UCD stellar masses range from $10^{6.3}$ to $10^{7.5}\,M_\odot$.}
    \label{fig:uiKs_mass_ssp}
\end{figure*}

We observe that the mass distributions for the three populations are quite different morphologically and cover different mass ranges.~At the low-mass end, this is due to differences in the respective selection functions.~The NGFS nuclei sample reaches lower masses than the CSS sample, because the spectroscopic selection function for the radial-velocity confirmation of CSSs in Fornax has a brighter cut-off \citep{Witt16} than the NGFS point-source detection limit (see Sect.~\ref{txt:sample}).~The mass distribution of dwarf spheroids is limited at the lower-mass end by the surface brightness detection limit of these systems \citep[see][]{eigenthaler2018}.~The nuclei population spans more than two orders of magnitude in stellar mass and shows a bimodal distribution, for which the peaks are located at $\log({\cal M}_\star/M_\odot)\!\simeq\!5.38$ and $\sim\!6.25$, the latter value is consistent with ultra-compact dwarf (UCD) masses \citep[e.g.,][]{misgeld11}.~Although the AIC marginally prefers one over two components, the Epanechnikov density distribution favors two components.~We find that the CSS distribution is formally trimodal\footnote{We point out that the AIC has a minimum at three components, but the AIC for two components is numerically very close to the formal minimum. In addition, the BIC gives two as the most likely number of components, which indicates that bimodality or trimodality are equally likely representations of the CSS mass distribution.} with its peaks located at $\log({\cal M}_\star/M_\odot)\!\simeq\!5.81, 6.02$ and $\sim\!6.35$, with the last component extending towards higher masses, reminiscent of UCDs as well.~We will not discuss this interesting feature of the Fornax globular cluster mass function in this work, as it requires an in-depth analysis of the sample selection function, but keep our focus on the dwarf galaxy spheroids and nuclei.~The dwarf spheroid mass distribution occupies a broad range of more than three magnitudes with a mild, but statistically non-significant bimodality ($\log[{\cal M}_\star/M_\odot]\!\simeq\!7.25$ and $\sim\!8.25$), with its single-Gaussian peak at $\log({\cal M}_\star/M_\odot)\!\simeq\!7.8$.~To our knowledge this is the first time such hints for multi-modalities in the nucleated spheroid mass and the nucleus mass distribution have been detected.~Together they may give us hints at the importance of different formation mechanisms of stellar nuclei.

\subsection{Stellar population properties}
\label{txt:spprops}

Color information in combination with population synthesis model predictions can be used to understand the stellar population properties of the nuclei, such as age, metallicity, and correlations with their mass.~The $u'i'K_s$ diagram helps minimize the age-metallicity degeneracy that affects broadband filters.~Figure~\ref{fig:uiKs_mass_ssp} illustrates the $u'i'K_s$ diagram with over-plotted SSP models from BC03, showing iso-metallicity tracks for the range $0.0001\!<\!Z\!<\!0.05$ with equivalent ages of $1, 2, 5,$ and $13.8$\,Gyr.~Filled circles stand for NGFS nuclei color-coded by their stellar masses.~As in the previous figures, radial-velocity confirmed CSSs are shown for comparison purposes.~We observe that the NGFS nuclei occupy approximately the bluer half of the CSSs $u'i'K_s$ spread, which is consistent with sub-solar metallicities ($Z\!<\!Z_{\odot}$) and/or a young stellar age component.~This is in line with the measurements of \cite{Paudel11}, who find from spectroscopy of relatively bright Virgo dwarf galaxies ($-18.5\!\ga\!M_{r,{\rm gal}}\!\ga\!-15.5$) that their nuclei ($-13.3\!\ga\!M_{r,{\rm nuc}}\!\ga\!-10.2$) cover a large range of metallicities from slightly super-solar ($+0.18$\,dex) to significantly sub-solar ($-1.22$\,dex) values.~However, it is still challenging to disentangle ages and metallicities for stellar systems older than a few Gyr at any metallicity based on photometry alone.~The inversion of broadband color information into stellar population parameters is notoriously difficult \citep{hansson12,powalka16b, powalka17} and is facing limitations in light of potentially as yet to be understood systematics related to the galaxy cluster environment \citep{powalka16a}.~Given the previous considerations, we refrain from assigning numerical age and metallicity values to the nuclei, but analyze them in groups.

\subsubsection{Bimodality in nucleus stellar population properties}
Although there is no clear mass-color relation, the mass bimodality of our NGFS nuclei is seen as mainly two groups in the $u'i'K_s$ plane, hereafter referred to as groups ${\cal A}$ and ${\cal B}$, and indicates a bimodality in their stellar population parameters.~The nuclei in group ${\cal A}$ have stellar masses $\ga\!10^{6}$\,M$_\odot$ and lie in the bluest color-color region of the $u'i'K_s$ plane with $(u'\!-\!i')_0\!<\!2.0$\,mag and $(i'\!-\!K_s)_0\!<\! -0.2$\,mag.~Nuclei in group ${\cal B}$ cover a more extended $u'i'K_s$ color space with redder average colors, i.e.,~$(i'\!-\!K_s)_0\!\ga\! -0.2$\,mag, and comprise objects with stellar masses $\la\!10^{6}$\,M$_\odot$\footnote{One exception in group ${\cal A}$ is the massive nucleus in the center of dwarf NGFS034050-354454 with a stellar mass of $\log({\cal M}_\star/M_\odot)\!\simeq\!7.26$ and colors consistent with $Z\!\simeq\!0.004$ and $\sim\!3$ Gyr (yellow symbol in Fig.~\ref{fig:uiKs_mass_ssp})}.~The mean masses for the groups are close to the masses of the two peaks in the nuclei stellar mass distribution, but it is worth noticing that from the total sample, 26 nuclei have high-quality $u'i'Ks$ photometry to robustly estimate their color-color distribution.~This is mainly the reason for the slightly different peaks between the bimodal mass distribution and the mean masses of the two groups considered here.~In any case, it is clear that according to the nucleus mass distribution from Figure~\ref{fig:massdistr}, there is a bimodality of nuclei which is also reflected in the $u'i'K_s$ diagram and that bimodality correlates with stellar population parameters specific for two nuclei sub-groups.

Assuming SSP-like stellar populations, we observe that our NGFS nuclei in group ${\cal A}$ host very metal-poor stellar populations ($Z\!<\!0.0004\!=\!0.02\,Z_{\odot}$) with luminosity-weighted ages older than $\sim\!2$\,Gyr.~In contrast, the nuclei in group ${\cal B}$ show colors equivalent to metallicities in the range $0.004\!<\!Z\!<\!0.02$ ($0.2\!<\!Z/Z_{\odot}\!<\!1$), and ages younger than $\sim\!2$\,Gyr.~The reddening vector in Figure~\ref{fig:uiKs_mass_ssp} illustrates how an intrinsic reddening of $A_V\!=\!0.5$\,mag, equivalent to $E_{B\!-\!V}\!=\!0.16$ for a Milky Way reddening curve, would affect the $u'i'K_s$ color-color space.~The reddening direction points towards increasing metallicity values, but does not affect the age significantly -- if anything, it pushes the colors towards older equivalent ages.~Alternatively, chemical abundance ratios different from the ones assumed in the solar-scaled BC03 models may bias the age and metallicity estimates.~Evidence that this is indeed the case comes from the study of \cite{paudel10} who found super-solar [$\alpha$/Fe] ratios in nuclei of Virgo dwarfs.~In their photometric study of compact stellar systems in the Virgo cluster, \cite{powalka16a} found intriguing offsets in multi-color relations pointing towards younger ages of some Virgo GCs.~Although test showed the influence of increased [$\alpha$/Fe] on colors, the team found that typical $\alpha$-element enhancements of Local Group GCs were producing too small color offsets to match the observations at old ages.~We, therefore, tentatively conclude that the younger ages of the nuclei in group ${\cal B}$ are not primarily due to intrinsic reddening and [$\alpha$/Fe] variations, but likely due to genuinely younger and more metal-rich stellar populations, which lie in terms of stellar mass in the low-mass mode of the nucleus stellar-mass bimodality.~Overall, these results point to different formation histories and perhaps different mechanisms of nucleus formation between the two groups.~We will come back to this in the discussion section.

\subsubsection{Comparison of nuclei with confirmed UCDs}
The nuclei masses are shown in color-code in the $u'i'K_s$ color-color diagram in Figure~\ref{fig:uiKs_mass_ssp} (top panel) and are compared to the corresponding stellar mass distribution of radial-velocity confirmed CSSs in Fornax (bottom panels).~From this CSS sample, UCDs are selected with a stellar-mass cut so that $\log{\cal M}_\star/M_\odot({\rm UCD})\!\geq\! 6.3$ \citep[e.g.][]{tay10, mie13}, avoiding any restriction in color (i.e.~metallicity), while GCs are selected with $\log{\cal M}_\star/M_\odot({\rm GC})\!<\! 6.3$ from the same parent CSS sample.~The final UCD sample exhibits stellar masses within the range $\log({\cal M}_\star/M_\odot)\!=\!6.3\!-\!7.4$, whereas the GCs cover the mass range $\log({\cal M}_\star/M_\odot)\!=\!5.4\!-\!6.3$, which is limited at the low-mass end by the spectroscopic selection function of the CSS sample \citep{Witt16}.~The majority of our NGFS nuclei are less massive than $\log({\cal M}_\star/M_\odot)\!\simeq\!6.8$ with the exception of three objects with a stellar mass of $\log({\cal M}_\star/M_\odot)\!\simeq\!7.2\!-\!7.3$ (see Fig.~\ref{fig:massdistr}), only one of which has a $u'i'K_s$ color and is plotted in Figure~\ref{fig:uiKs_mass_ssp}.

Comparing the stellar masses and stellar population parameters of nuclei with those of UCDs reveals that the members of the low-mass mode of our sample nuclei cannot be the progenitors of Fornax UCDs.~These dwarf nuclei have simply too low masses to be considered a parental population.~However, nuclei that are members of the high-mass mode could potentially be considered progenitors of metal-poor Fornax UCDs.~Considering that the initial mass of UCD progenitors may be even higher than their current mass and given their likely mass evolution, as suggested in some UCD formation scenarios that involve stripping \citep[e.g.][]{zin88, bas94, bek01, bek03, goe08, pfe13}, even this evolutionary connection may be questionable given the similar present-day masses of high-mass dwarf nuclei and metal-poor UCDs.~This may not be the case for the most massive dwarf nucleus in our sample, NGFS034050-354454n, which qualifies as a potential intermediate-metallicity Fornax UCD seed, even after 90\% of its present-day mass is lost.~In any case, our data suggest that the progenitors of the massive, metal-rich UCDs in Fornax have long been destroyed and have no present-day counterparts in dwarf galaxy nuclei.~This is consistent with previous spectroscopic studies \citep{evs07, fra12}.

\subsubsection{Comparison of nuclei with confirmed GCs}
The picture is different when we compare the properties of nuclei and GCs.~Essentially all of our nuclei can be considered potential future GCs, once the spheroid envelopes surrounding them are stripped during their dynamical evolution within the Fornax cluster \citep{goe08, bek03, smi13, smi15}.~Such potential nuclei remnants may become future members of the Fornax GC system.~The younger stellar ages of the low-mass mode nuclei indicate extended star formation histories and, therefore, prolonged chemical enrichment processes that may lead to signficant abundance spreads, especially in Fe-peak elements.~Such future GCs could be clearly identified with the next-generation of 30m class telescopes via their spreads in stellar iron abundances, something that has been measured in numerous Milky Way star clusters \citep[e.g.,][]{wil12}. Alternatively, high-spatial resolution imaging allows one to identify remnant nuclei candidates in the half-light radius vs.~luminosity parameter space, which has been done for the Local Group star clusters \citep{ma06}.~Furthermore, the characteristic age-metallicity relation found for a subset of Galactic GCs \citep{van13, lea13} is consistent with our observation of decreasing stellar ages in more metal-rich nuclei.~This suggests that at least the Galactic GC sub-population with a significant age-metallicity relation could have in part their origins in the cores of nucleated dwarf galaxies \citep[see also][]{mar09, for10, dot11,deBoer15}.

\subsection{Differences between nuclei and galaxy spheroids}
\begin{figure}[t!]
    %trim=left bottom right top
    \centering
     \includegraphics[trim=0.8cm 0.6cm 0.2cm 1.cm,clip,width=\columnwidth]{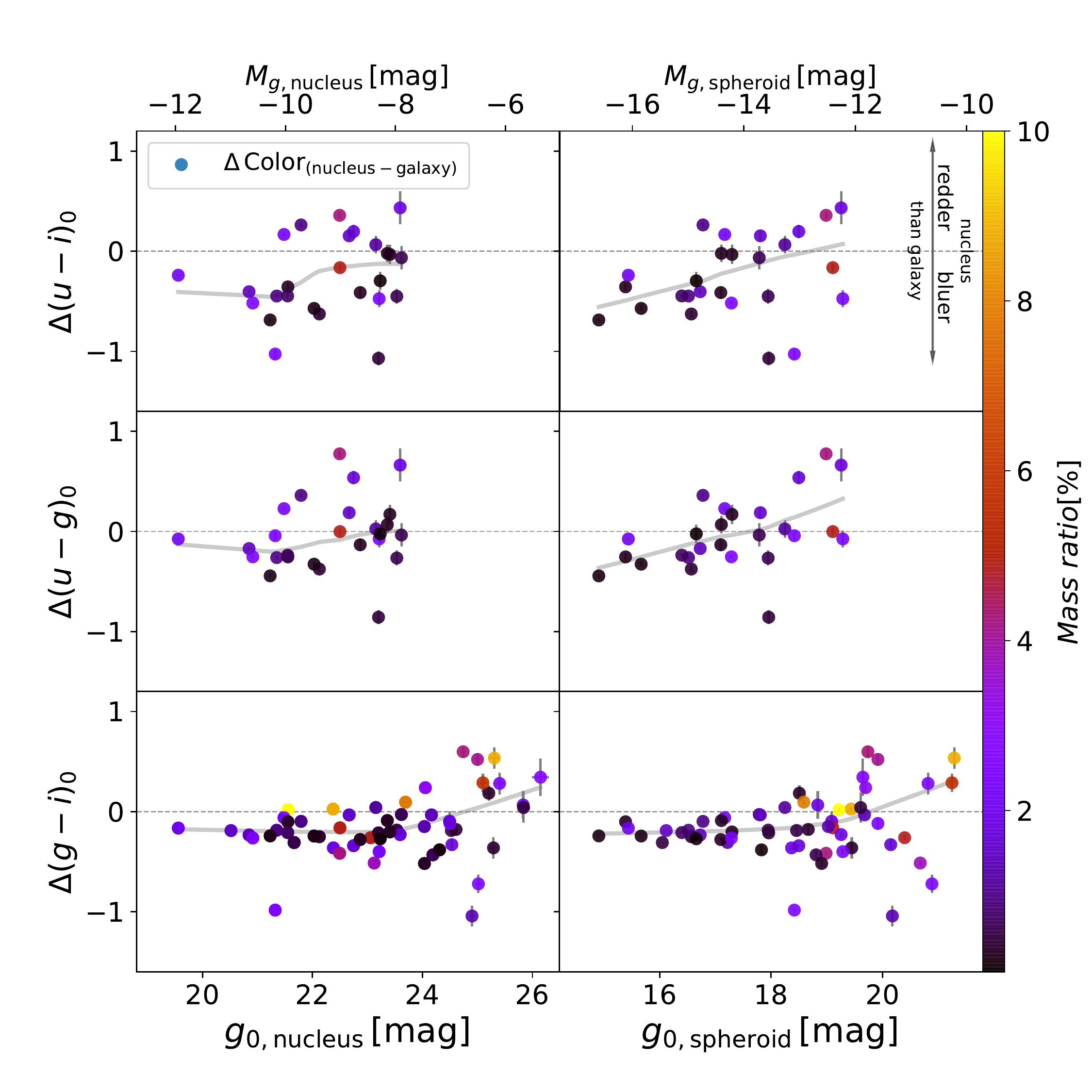}
     \caption{Color differences between the nucleus and its host dwarf galaxy spheroid in $\Delta(u'\!-\!i')_0$, $\Delta(u'\!-\!g')_0$, and $\Delta(g'\!-\!i')_0$ (top to bottom panels) vs.~$g_{0,{\rm nucleus}}$ (right panels) and vs.~$g_{0,{\rm spheroid}}$ (left panels).~Color code of the symbols shows the corresponding nucleus-to-galaxy mass ratio encoded by the vertical colorbar.~The gray curves represent LOWESS fits to the data.~The arrows in the top panel show the directions in which the nucleus becomes redder or bluer than its host galaxy.~Note that photometric errors are for most data points smaller than the symbol size.}
   \label{fig:cdiff}
\end{figure}

The color difference in various filter combinations between the nucleus and its host galaxy spheroid is shown in Figure~\ref{fig:cdiff} as a function of nucleus luminosity ($g_{0,{\rm nuc}}$) and galaxy spheroid luminosity ($g_{0,{\rm sph}}$).~The nucleus-to-galaxy mass ratio is encoded by the symbol color and has a range from $0.1\%-10\%$.~The color differences are more correlated with the spheroid light than with the nucleus luminosity, which implies that the mechanisms that lead to the color offsets must be acting more on galaxy scales rather than nucleus scales.~One can think of processes that are correlated with the total dwarf galaxy mass which, for instance, lead to more massive galaxies holding on more efficiently and longer to their gas supply than their small-mass counterparts; this is especially true in the galaxy cluster environment.

We find relations for the color differences $\Delta(u'\!-\!i')_0$ and $\Delta(u'\!-\!g')_0$ vs.~$g_{0,{\rm nuc}}$ and $g_{0,{\rm sph}}$.~These near-UV+optical colors map a broader SED range which is more sensitive to changes in stellar population parameters, while a shallower trend in $\Delta(g'\!-\!i')_0$ is consistent with the narrower SED coverage.~This is mainly due to the enhanced sensitivity of the u-band to the Balmer break flux compared to the redder filters.~Bright nuclei ($M_{g,{\rm nuc}}\!\lesssim\!-10$), on average, $\Delta(u'\!-\!i')_0\!=\!-0.35$, $\Delta(u'\!-\!g')_0\!=\!-0.15$, and $\Delta(g'\!-\!i')_0\!=\!-0.2$\,mag bluer than their host galaxy, indicative of younger and/or more metal-poor stellar populations (see Fig.~\ref{fig:uiKs_mass_ssp}).~According to the BC03 models, these color differences correspond consistently to an age difference of $\Delta t/t\simeq\!-0.8$ at old absolute ages ($\sim\!13$ Gyr) and low metallicities $Z\!=\!0.0001\!-\!0.004$ ([Fe/H]~$\!\simeq\!-2.3$ to $-1.6$\,dex), and an age difference of $\Delta t/t\simeq\!-0.5$ at young absolute ages ($\sim\!2$ Gyr) and solar metallicities.~It can also be attributed to a metallicity difference of $\Delta {\rm [Fe/H]}\!\approx\!-1$\,dex at old absolute ages ($\sim\!13$ Gyr) and intermediate-to-low metallicities ([Fe/H]~$\!\simeq\!-0.5$ to $-1.6$\,dex) and to $\Delta {\rm [Fe/H]}\!\approx\!-0.3$\,dex at solar metallicities, respectively.~These metallicity differences change at young absolute ages ($\sim\!2$ Gyr) to $\Delta {\rm [Fe/H]}\!\approx\!-1.3$\,dex at intermediate to low metallicities and $\Delta {\rm [Fe/H]}\!\approx\!-0.6$\,dex at solar metallicities, respectively.~Similar color differences have been reported in previous studies \citep[see][]{Lotz04, Cote06, Turner12}.~However, owing the depth of the present data, we find that there is a transition point around $g_{0,{\rm nuc}}\!\approx\!24$\,mag ($M_{g,{\rm nuc}}\!\approx\!-7.5$) or $g_{0,{\rm sph}}\!\approx\!19$\,mag ($M_{g,{\rm sph}}\!\approx\!-12.5$), where the average $(g'\!-\!i')_0$ offset becomes insignificant at the expense of an increasing galaxy-to-nucleus color variance.~This feature is noticeable in $(u'\!-\!i')_0$ and $(u'\!-\!g')_0$ colors already at brighter nucleus luminosities around $g_{0,{\rm nuc}}\!\approx\!22$\,mag ($M_{g,{\rm nuc}}\!\approx\!-9.5$) or $g_{0,{\rm sph}}\!\approx\!17.5$\,mag ($M_{g,{\rm sph}}\!\approx\!-14$).

Another important feature is the steep relation seen in the $\Delta(u'\!-\!i')_0$ and $\Delta(u'\!-\!g')_0$ colors vs.~$g_{0,{\rm sph}}$: here we observe that low nucleus-to-galaxy mass ratios ($<2\%$) only occur up to a magnitude $g_{0,{\rm sph}}\!\approx\!17.5$\,mag ($M_{g,{\rm sph}}\!\approx\!-14$) where we find almost exclusively blue nuclei, while for fainter dwarfs we have a mixture of nucleus-galaxy color differences and nucleus-to-galaxy mass ratios.~We note that all the trends described here are not due to photometric uncertainties, but stellar population properties that vary substantially from nucleus to nucleus.

These results paint the following picture: as a dwarf nucleus begins to grow, starting with a low nucleus-to-galaxy mass ratio, its stellar population content is dominated by more metal-poor and/or younger stars than the typical star in its host spheroid.~Nuclei with higher nucleus-to-galaxy mass ratios must have either reached higher metal enrichment at similar ages or were formed earlier with enough time for their stellar populations to evolve and redden sufficiently.~A distinction between these two scenarios could easily be made using spectroscopically determined [$\alpha$/Fe] ratios, which are indicators of star formation timescales \citep[e.g.,][]{mat86}, allowing us to discern between prolonged star formation histories vs.~short and early star-formation bursts \citep[e.g.,][]{tsu95, mat01}, which may be driven by the environment \citep[e.g.,][]{tho05, Puzia05}.

\begin{deluxetable*}{lcccc}[!t]
\tabletypesize{\scriptsize}
\tablecaption{Scaling relations \label{tab:scal_rel}}
\tablehead{
\colhead{Fit}            & \colhead{Parameters} & \colhead{Fornax}      & \colhead{NGFS dwarfs + ETGs} & \colhead{NGFS dwarfs + LTGs} }
\startdata
                         &                      &   ${\cal M}_{\rm nucleus}\,{\rm vs.}\,{\cal M}_{\rm galaxy}$  &    &                \vspace{1mm}\\
\hline 
Linear regression        & $a$                  &  $0.766 \pm 0.048$    & $0.723 \pm 0.046$          &  $0.480 \pm 0.033$     \\   
$y=ax+b$                 & $b$                  &  $-0.096 \pm 0.396$   & $0.197 \pm 0.392$          &  $1.850 \pm 0.305$      \\
\hline
Polynomial-fit degree=3  & $a$                  &  $0.068 \pm 0.038$    & $0.055  \pm 0.038$       & $0.031   \pm 0.025$      \\
$y=ax^3+bx^2+cx+d$       & $b$                  &  $-1.573 \pm 0.946$   & $-1.273 \pm 0.976$       & $-0.703  \pm 0.654$      \\
                         & $c$                  &  $12.418 \pm 7.881$   & $10.245 \pm 8.214$       & $5.646   \pm 5.685$      \\
                         & $d$                  &  $-28.085 \pm 21.677$ & $-22.923 \pm 22.821$     & $-10.212 \pm 16.313$  \vspace{1mm}   \\
%with \,\, $y=\log({\cal M}_{\rm nucleus})$   & and \,\,\,$x=\log({\cal M}_{\rm galaxy})$    &                      &                          &     \vspace{1mm}   \\
\hline
                        &                      &   $\eta_n={\cal M}_{\rm nucleus}/{\cal M}_{\rm galaxy}\,{\rm vs.}\,{\cal M}_{\rm galaxy}$       &    &               \vspace{1mm} \\
\hline
Linear regression       & $a$                  &  $-0.234 \pm 0.048$     & $-0.277 \pm 0.046$     &  $-0.520 \pm 0.033$     \\   
$y=ax+b$                  & $b$                  &  $-0.096 \pm 0.396$     & $0.198 \pm 0.392$      &  $1.850  \pm 0.305$     \\
\hline
Polynomial-fit degree=3 & $a$                  &  $0.069  \pm 0.038$   & $0.055 \pm 0.038$       & $0.031 \pm 0.025$      \\
$y=ax^3+bx^2+cx+d$        & $b$                &  $-1.573 \pm 0.946$   & $-1.273 \pm 0.976$       & $-0.703 \pm 0.654$    \\
                        & $c$                  &  $11.418 \pm 7.881$   & $9.245 \pm 8.214$       & $4.646 \pm 5.685$      \\
                        & $d$                  &  $-28.087 \pm 21.677$ & $-22.924 \pm 22.821$     & $-10.212 \pm 16.313$    \vspace{1mm} \\  
\enddata
\tablecomments{For the nucleus vs.~galaxy mass relation (${\cal M}_{\rm nucleus}\,{\rm vs.}\,{\cal M}_{\rm galaxy}$) we set $y\!\equiv\!\log({\cal M}_{\rm nucleus})$ and $x\!\equiv\!\log({\cal M}_{\rm galaxy})$.~For the mass ratio ($\eta_n$) relation as a function of galaxy mass ($\eta_n\,{\rm vs.}\,{\cal M}_{\rm galaxy}$) we define $y\!\equiv\!\log(\eta_n)$ and $x\!\equiv\!\log({\cal M}_{\rm galaxy})$.}
\end{deluxetable*}

\section{Discussion}
\subsection{Formation mechanisms}
The astrophysical mechanisms responsible for the differences in stellar population content of the galaxy spheroid and the nucleus are numerous, but can be categorized to be mainly due to two processes: {\it i)} the inflow of gas into the nuclear regions which triggers star formation processes \citep[e.g.,][]{vdBergh86, Antonini15}, and {\it ii)} the accretion of GCs into the galaxy central regions via dynamical friction \citep[e.g.,][]{Tremaine75, Lotz01}.~However, the relatively shallow dwarf galaxy potentials can be easily affected by environmental and secular processes.~Secular processes such as stellar winds, supernova (SN) and black-hole (BH) feedback can affect the nucleus formation and evolution, for instance, by helping with gas supply to the nuclear regions through stellar winds from newly formed stars \citep[radiation drag,][]{kawa_ume02} or slowing down the nucleus growth due to SN-driven winds contributing to the mass loss in dwarf galaxies and likely changing the dynamical friction timescales for orbiting star clusters to sink to the center \citep{Lotz01}, or dynamically heating the nuclear cluster due to a massive central BH \citep{Antonini15}.~Whether the gas comes from disk instabilities, galaxy mergers (with some gas content) or primordial gas, the dynamical timescales for the gas to sink down to the nuclear reservoir depend on the size of the galaxy, being longer with increasing galaxy size and, thus, mass \citep{eigenthaler2018}.~Therefore, in more massive galaxies the inflowing material would have more time to fragment and undergo star-formation, leaving smoother and relatively steeper stellar population gradients imprinted in the spheroid component.~Consequently, the spheroids of low-mass dwarfs would have smaller and more stochastic population gradients due to gas and/or GCs having shorter sink-in timescales, leading to a more stochastic color difference between nucleus and host galaxy than for more massive dwarfs.~This is exactly what we observe in Figure~\ref{fig:cdiff} for $M_{g,{\rm sph}}\le-14$ and $M_{g,{\rm nuc}}\le-9.5$\,mag.

\begin{figure}[t!]
    %trim=left bottom right top
	 \includegraphics[trim=0.2cm 1.3cm 0.8cm 3.cm,clip,width=0.49\textwidth]{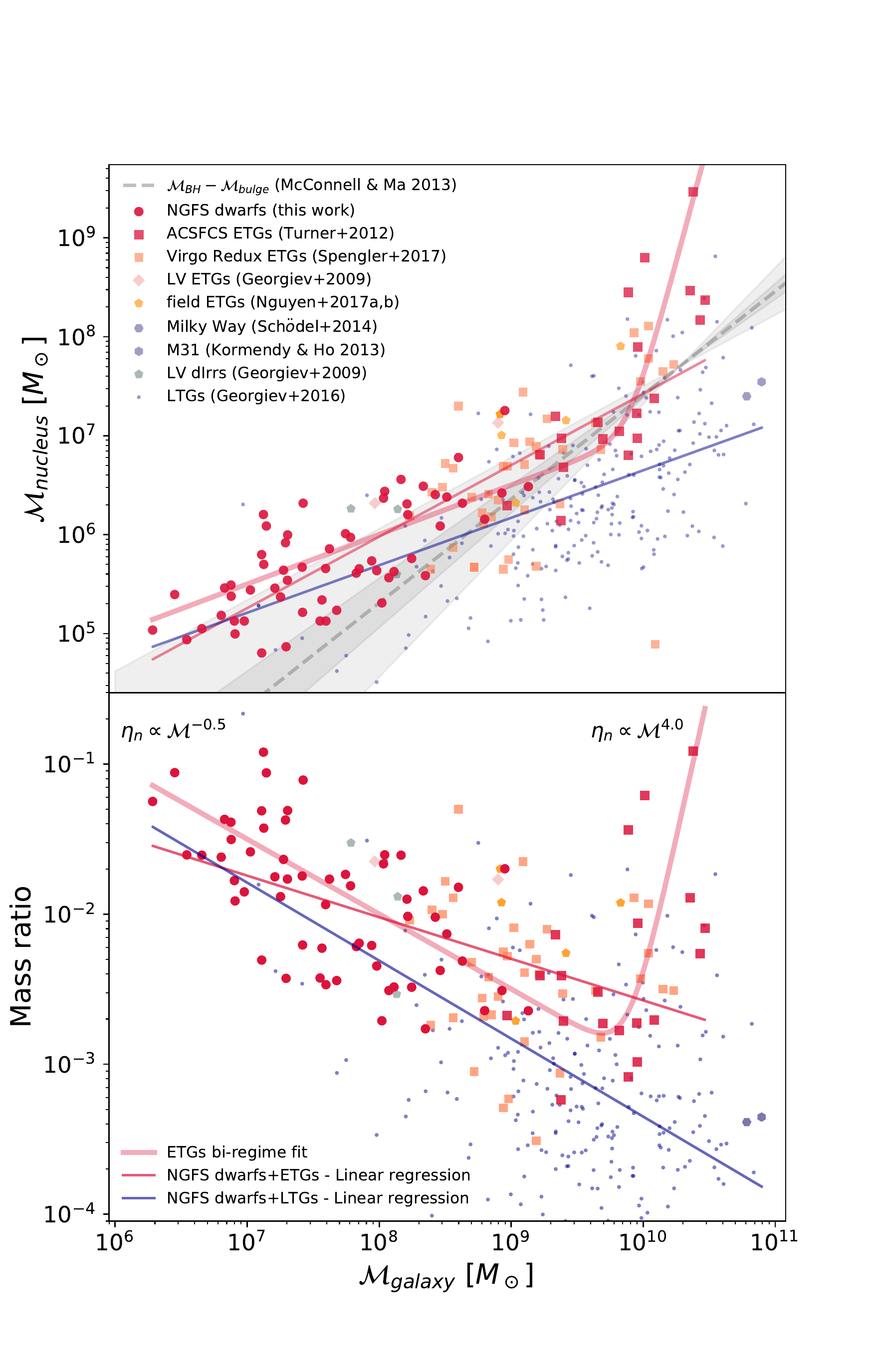}
    \caption{{\it Top panel:} Nucleus vs.~host galaxy stellar mass for different subsets of nucleated NGFS galaxies and other galaxies from the literature.~The SMBH mass vs.~galaxy mass relation from \cite{mcconell13} is shown with $1\sigma$ and $3\sigma$ limits as dashed-line and shaded regions, respectively.~{\it Bottom panel:} Nucleus-to-galaxy mass ratio ($\eta_n\!=\!{\cal M}_{\rm nuc}/{\cal M}_{\rm gal}$) vs.~galaxy mass.~We approximate the ETG data with a bi-regime fit for which the slopes of the low-mass and high-mass galaxy masses are labeled.} 
\label{fig:scaling_relations}
\end{figure}

\subsection{Scaling relations}
One of the physical scaling relations that nuclei follow is the nucleus-to-galaxy mass relation (see Fig.~\ref{fig:scaling_relations}) which has been shown to hold for bright ETGs \cite[e.g.,][]{Spengler17}.~To test whether such a relation applies to our dwarf galaxy sample we use the masses derived for the nuclei in this work (Sect.~\ref{txt:stellar_mass}) and their spheroid masses from \cite{eigenthaler2018}.~In order to compare our sample with nucleated galaxies at higher masses and those located in different environments, we make use of literature data.~For the bright nucleated galaxies in the Fornax cluster sample from ACSFCS \citep{Turner12}, the corresponding nucleus masses are estimated with the same method described in Section~\ref{txt:stellar_mass} using their $g,z$ photometry.~For their host galaxies the $B,V$ magnitudes were obtained from HyperLEDA and together with the relations from \cite{Bell03} we estimate the masses in a self-consistent procedure as for the NGFS dwarf galaxies \citep{eigenthaler2018}.~For the Virgo cluster, we use the masses estimated in the Virgo Redux work \citep{Spengler17} for nuclei and hosts.~In addition, we include the results of recent studies by \cite{Nguyen17a, Nguyen17b} for field ETGs to estimate the dynamical mass for their central BH and their nuclear star cluster.~In \cite{Nguyen17a}, the central SMBH dynamical mass for NGC\,404 was estimated to be $1.5\times10^5M_\odot$.~In \cite{Nguyen17b} four field ETGs were studied.~Three of them were found to contain BHs with masses of $2.5\times10^6M_\odot$ (M32), $8.8\times10^5M_\odot$ (NGC\,5102) and $4.7\times10^5M_\odot$ (NGC\,5206).~The catalog of nucleated late-type galaxies (LTGs) comes from \cite{Georgiev16} with 228 moderately inclined spiral galaxies with morphological type code $T\!\geq\!3$ or later than Sb at distances $<\!40$\,Mpc.~Nuclei from Local Volume (LV) dwarf irregular (dIrr) and early-type dwarf galaxies come from \cite{Georgiev09}.~For the Milky Way and Andromeda nuclear star clusters we use ${\cal M}_{\rm MW,NSC}\!=\!(2.5\pm0.4)\times10^7M_\odot$ from \cite{schodel14} and ${\cal M}_{\rm M31,NSC}\!=\!(3.5\pm0.8)\times10^7M_\odot$ from \cite{kormendy13}.

The relation between nucleus and host galaxy stellar mass is shown in Figure~\ref{fig:scaling_relations} (top panel), where we see a clear mass correlation between nuclei and their host galaxies across the entire galaxy mass range ($6\!\la\!\log[{\cal M_*}/M_\odot]\!\la\!11$).~For ETGs, this relation is shallower for lower-mass galaxies compared to the massive galaxy regime with a break in slope around $\log({\cal M_*}/M_\odot)\!\simeq\!9.7$.~The relation for dwarf galaxies scales as $\eta_n\!\propto\!{\cal M}_{\rm gal}^{-0.5}$, while for massive ellipticals it follows a much steeper relation, $\eta_n\!\propto\!{\cal M}_{\rm gal}^{4}$.~The discussion of the astrophysical reasons for these two regimes go far beyond the scope of this paper.

Compared to ETGs, there is a higher dispersion and a less inclined slope for the LTG relation, which is more noticeable in the massive galaxy range, as was already pointed out by \cite{Georgiev16}.~This indicates that the nuclei in LTGs are on average less massive at a fixed host galaxy mass than nuclei in ETGs.~The weighted linear and polynomial least-square fits are shown in Table~\ref{tab:scal_rel}.~Based on considerations that the mechanisms, that are responsible for the build-up of the central massive objects are similar for nuclei and massive BHs, previous studies have discussed their possible evolutionary connection \citep[e.g.,][]{fer06, neu12}.~For the purpose of comparison, we illustrate the black-hole-galaxy mass relation from \cite{mcconell13} with its $1\sigma$ and $3\sigma$ uncertainty limits.~This relation was obtained from fitting the compilation of 35 ETGs with dynamical measurements of the bulge stellar mass, with mass range of $10^{9}\!-\!10^{12}M_\odot$.~Their sample is well populated for bulges more massive than $2\times10^{10}M_\odot$ (see their Figure~3).~Assuming that we can extrapolate to lower bulge masses, we see in Figure~\ref{fig:scaling_relations} that the scaling relation for BHs and their ETG hosts is similar to the relation between nuclei and their host galaxy mass down to $10^9M_\odot$.~For galaxies with lower masses the nucleus-to-galaxy mass relation becomes flatter.

The mass ratio between nucleus and its host galaxy ($\eta_n\!=\!{\cal M}_{\rm nuc}/{\cal M}_{\rm gal}$) as a function of galaxy mass is shown at the bottom panel of Figure~\ref{fig:scaling_relations}.~We find a clear anti-correlation of $\eta_n$ vs.~${\cal M}_{\rm gal}$ over the entire galaxy mass range, i.e.~the lower the galaxy mass the more significant becomes the nucleus.~In the massive galaxy regime (${\cal M}_{\rm gal}>10^9\,M_\odot$) we note a large scatter in the relation, which is the main reason why previous studies assumed a constant equivalent luminosity ratio $\eta_{n,L}\!=\!{\cal L}_{\rm nuc}/{\cal L}_{\rm gal}$ for their samples, like for instance, the ACS Virgo nucleated galaxies with $\langle\eta_{n,L}\rangle\!=\!0.30\% \pm0.04\% $ \citep{Cote06}, $\langle\eta_{n,L}\rangle = 0.41\% \pm0.04\% $ for the ACS Fornax nucleated galaxies \citep{Turner12} and $\langle\eta_{n,L}\rangle = 0.1\%$ for LTGs \citep{Georgiev16}.~However, in the faint dwarf galaxy regime (${\cal M}_{\rm gal}\leq10^9\,M_\odot$) there appears a clear and strong trend, reaching to $\eta_n\!\simeq\!10\%$ for a dwarf galaxy with a stellar mass of $10^7M_\odot$.~The four low-mass ETGs studied by \cite{Nguyen17a,Nguyen17b} have $\eta_n$ values up to $1.7\%$, which are in agreement with the general trend.~The extension of the $\eta_n$ vs.~${\cal M}_{\rm gal}$ anti-correlation towards the faint dwarf galaxy population appears to be similar for ETGs and LTGs.~This, in turn, suggests that nuclei at the smallest masses are subject to localized processes that work on parsec scales within the galaxy core regions independent of galaxy type.

\subsection{Comparison with theoretical predictions}
The two proposed formation scenarios for nuclei are globular cluster infall due to dynamical friction \citep{Tremaine75} and in situ star formation \citep{vdBergh86}.~The latter needs a mechanism to funnel gas into the galaxy center.~Some studies suggest mechanism to carry the gas inwards to be galaxy mergers between two disk galaxies \citep{Mihos94}, supernova feedback outflows that become stalled because the intergalactic medium (IGM) pressure prevents the gas from escaping the dwarf galaxy \citep{Babul92}, and gas disks embedded in an old stellar spheroid \citep{Bekki07}.~Observational studies for ETGs have argued that the predominant mechanism for nucleus formation in more massive galaxies are dissipative processes, sinking gas to the central galaxy regions with star formation occurring {\it in situ}, while for low-mass galaxies nucleus formation occurs via GC infall due to short dynamical timescales \citep[e.g.][]{Lotz04, Cote06, Turner12}.~In this context, we note that in the Virgo cluster, more than 50\% of the bright early-type dwarfs were found to show underlying disk features, with the disk fraction decreasing to only a few \% for such dwarfs fainter than $M_B\!=-\!15.5$\,mag \citep{Lisker06a}, corresponding to $\log({\cal M}_\star/M_\odot)\simeq8.6$ \citep[see Fig.~7 in][]{eigenthaler2018}.~Moreover, about 15\% of the Virgo early-type dwarfs brighter than $M_B\!=\!-15.5$\,mag reveal blue centers, which were spectroscopically shown to correspond to recent star formation \citep{Lisker06b}.~In a more recent work, \cite{Spengler17} have compared their multi-band photometry of nuclei with scaling relation predictions from \cite{Bekki07} and \cite{Antonini15} and inferred that there is no single preferred formation scenario for nuclei, suggesting a mix of processes instead \citep[see also][]{dar11}.~We proceed to compare the nucleus sample available for the Fornax cluster (NGFS and ACSFCS nuclei) with scaling relation predictions for different formation scenarios in a similar approach as in \cite{Spengler17}.~Figure~\ref{fig:theory_scaling_relations} illustrates the comparison of empirical results with theoretical predictions.

\subsubsection{\cite{Bekki07} model predictions}
\begin{figure}[t!]
    %trim=left bottom right top
	 \includegraphics[trim=0cm 1.0cm 1.5cm 3.0cm,clip,width=1.0\columnwidth]{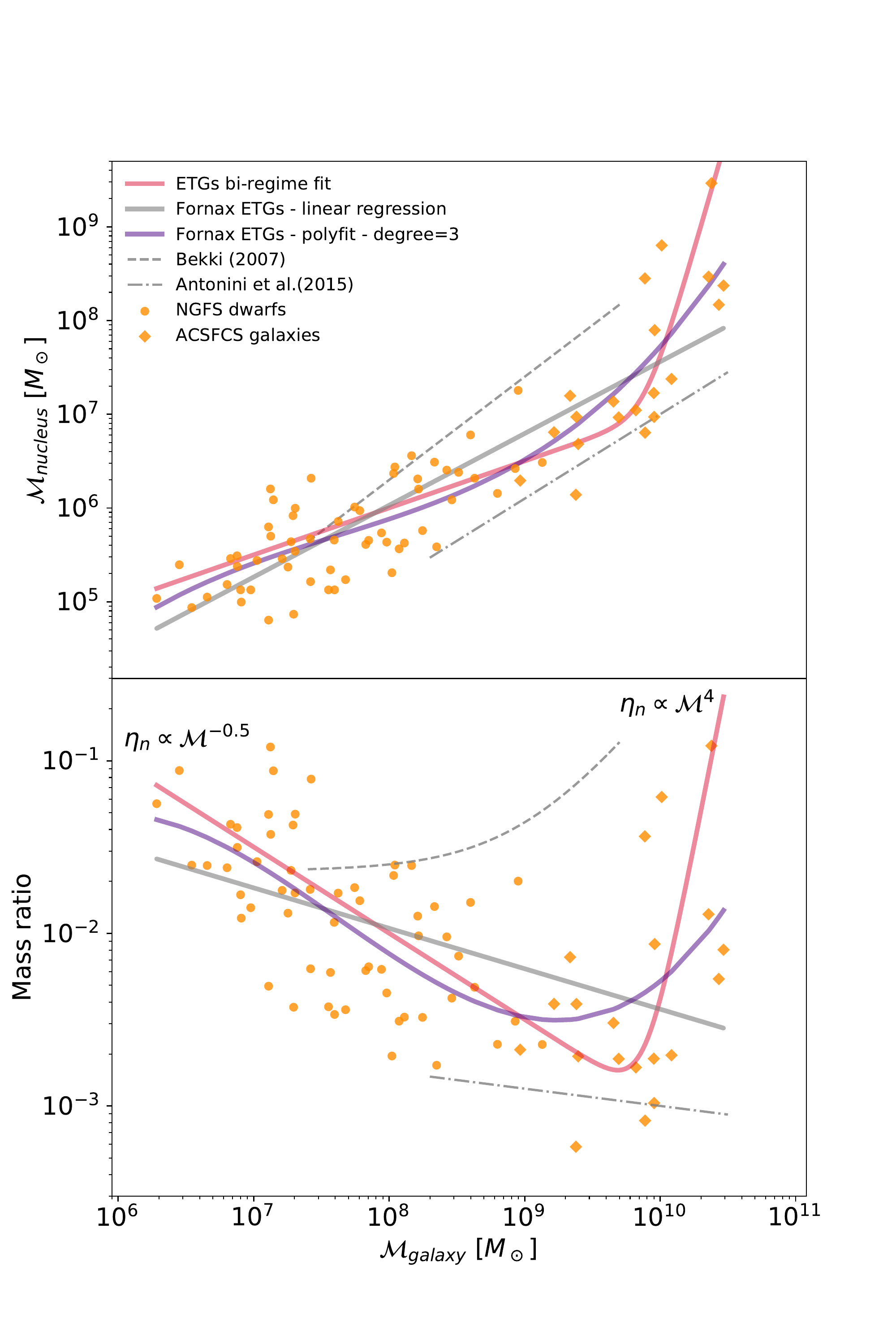}
    \caption{Scaling relations of the nucleus and galaxy masses.~{\it Top panel}: Nucleus vs.~galaxy mass relation for all nucleated galaxies in the Fornax core region (NGFS dwarfs and ACSFCS sample).~{\it Bottom panel}:~Nucleus-to-galaxy mass ratio ($\eta_n\!=\!{\cal M}_{\rm nuc}/{\cal M}_{\rm gal}$) as a function of galaxy mass.~Solid lines show the weighted least-square fits, the numerical values of which are shown in Table~\ref{tab:scal_rel}.~Both panels show model predictions for nucleus formation.~See the legend and text for more details.} 
    \label{fig:theory_scaling_relations}
\end{figure}

Pure dissipative models such as the one put forward by \cite{Bekki07}, which takes into account feedback from SNe and super-massive BHs, depend mainly on the spheroid mass ($0.025<{\cal M}_{\rm sph}/10^9 M_\odot<5.0$), the initial gas mass fraction ($0.02\leq f_{\rm gas}\leq0.5$), the spheroid surface brightness (SB), and the chosen IMF (bottom or top-heavy).~Some of the more relevant nucleus properties in the numerical results of this model are that $\eta_n$ can reach up to 5\%, more massive spheroids have more metal-rich nuclei and less massive spheroids can hold a young nucleus due to longer timescales of nucleus formation.~Compared to the Fornax nucleated galaxy sample, the \citeauthor{Bekki07} model (gray-dashed curves in Fig.~\ref{fig:theory_scaling_relations}) reproduces the $\eta_n$ values for galaxies with masses of ${\cal M}_{\rm gal}\lesssim10^8\,M_\odot$, but predicts too massive nuclei in more massive galaxies relative to the observations.~The predicted trend in the $\eta_n$-galaxy mass relation is too steep for massive nucleated galaxies in contrast to the empirical results.~Although the observed mass ratios reach up to 10\%, they do so only at the lowest ($\sim\!10^7\,M_\odot$) and highest sampled masses ($\ga\!10^{10}\!M_\odot$), while the model reaches those values at smaller masses.~For masses of the order of $10^{9.5}\,M_\odot$, the mass ratio $\eta_n$ for our sample is about one order of magnitude smaller than the predicted values.~Clearly the theoretical ingredients of the \cite{Bekki07} model ought to be adjusted to better reproduce the observed hockey-stick trend of the $\eta_n-{\cal M}_{\rm gal}$ relation, where stronger suppression within the model framework of the nucleus mass accumulation process at intermediate masses ($10^{8}\!-\!10^{10}\,M_\odot$) seems necessary.~In light of the relatively high fraction of disk components in intermediate-mass dwarfs \citep[see][]{Lisker06a}, this may be accomplished by either an enhanced disk/spheroid growth mode and/or suppression of the nuclear mass accumulation mechanism, e.g.~via advective angular momentum transport, bar instabilities, and/or the presence of a central black hole \citep{curir08, curir10, foyle10, goz15, james18}.

\begin{figure*}[t!]
    %trim=left bottom right top
    \centering
    \includegraphics[width=0.95\textwidth]{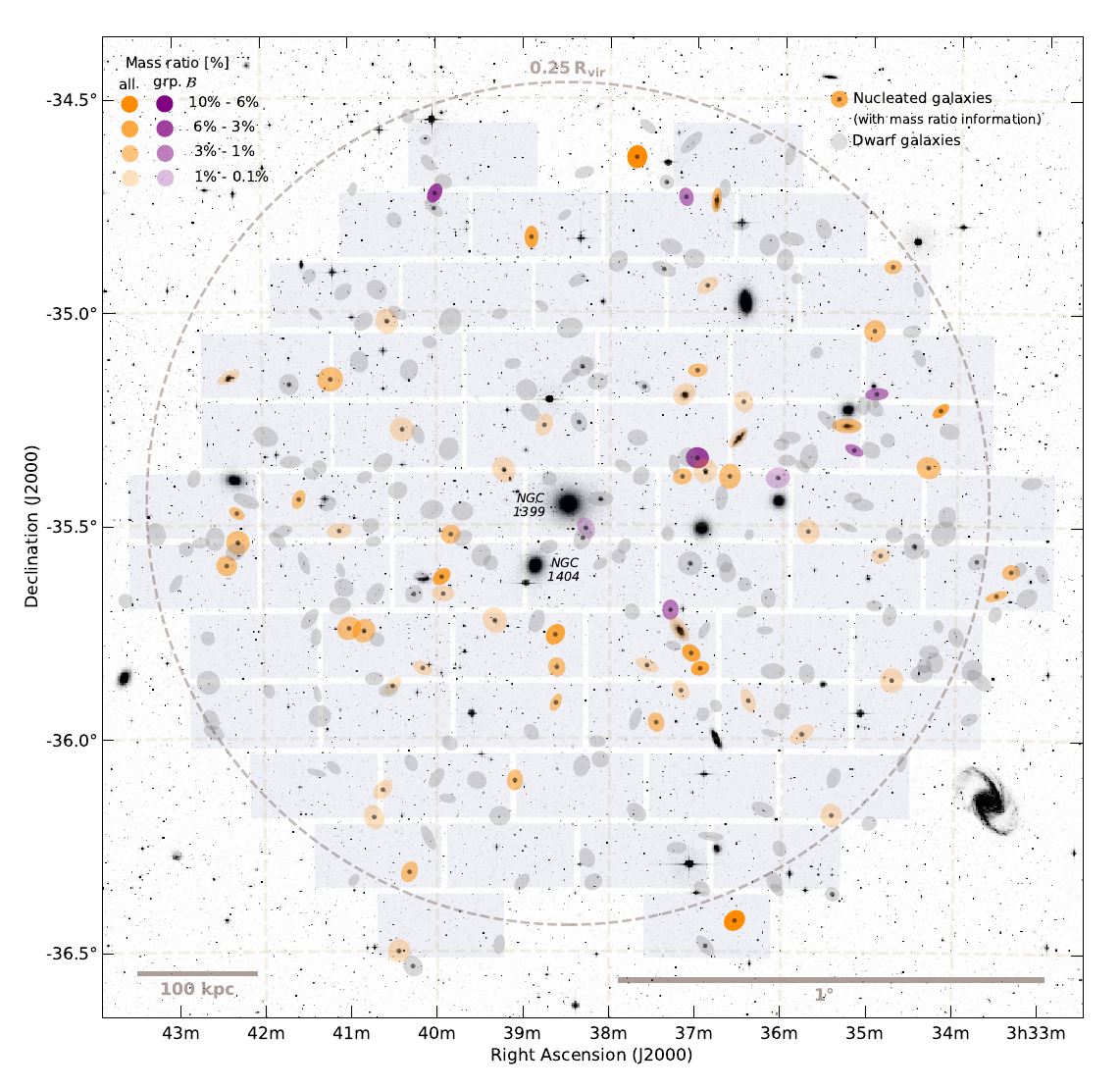}
    \caption{Illustration of the central region of the Fornax galaxy cluster showing the spatial distribution of non-nucleated (gray symbols) together with nucleated dwarfs, which are shown as orange symbols (NGFS dEN and ACSFCS) if mass ratio information is available.~Otherwise, they are shown in gray scale as well.~The symbol transparency parameterizes the mass ratio as $\eta_n\!=\!{\cal M}_{\rm nuc}/{\cal M}_{\rm gal}$ ranging from 10\%-0.1\% of the host galaxy mass, which is indicated in the top left corner.~Group ${\cal B}$ nuclei (see Sect~\ref{txt:spprops}) which are younger and more metal-rich than group ${\cal A}$ are shown in purple.} 
    \label{fig:spatial_mr}
\end{figure*}

\vspace{0.2cm}
\subsubsection{\cite{Antonini15} model predictions}
An hybrid approach to modeling the formation of galaxy nuclei was undertaken by \cite{Antonini15} where two nucleus formation models are considered.~The first model is the cluster-inspiral (CliN) model, which simulates star cluster mergers in the center of an isolated galaxy with a pre-existing central BH.~The second is a galaxy formation (GxeV) model, which tracks the evolution of baryonic structures in dark-matter merger trees.~The GxeV model includes galaxy evolution, dissipative processes, mergers between galaxies, tidal interactions, and coexistence with super-massive BHs.~This hybrid approach considers two possibilities for nucleus growth from high-redshift to the present-day by migration of star clusters and/or in situ formation.~Both models, CliN and GxeV, have similar scaling relations but GxeV has a larger dispersion in the nucleus masses than CliN does.~In addition, CliN cannot form nuclei more massive than a few $10^7\,M_\odot$, but GxeV can.~\citeauthor{Antonini15} tested the case without BH heating and found that both models still are able to form massive nuclei without any constraint on velocity dispersion or galaxy mass.~One interesting prediction is that nuclei can be formed with one mechanism alone, the in situ star formation.~However, a shallower slope is then obtained for the nucleus-galaxy mass relation relative to the scenario when both mechanisms are at work.

When comparing the predictions (CliN and GxeV are similar in this parameter space, dash-dotted lines in Fig.~\ref{fig:theory_scaling_relations}) with Fornax nucleated galaxies, the predicted masses tend to be a factor of a few smaller than the observed ones over the mass range covered by the models ranging from $2\times10^8\,M_\odot$ to $3\times10^{10}\,M_\odot$.~\citeauthor{Antonini15}~notice the offset and argue that the underweight of model nuclei is due to the interaction of the nucleus with the central massive BH, which makes the nucleus lose stars faster, in addition to galaxy mergers, where BH binaries form and efficiently eject surrounding stars.~These effects have a greater impact in more massive galaxies than in low-mass dwarfs.~The overall prediction by \cite{Antonini15} is that both mechanisms are likely active during nucleus growth and that their relative contribution depends on the star-cluster formation efficiency.~These models show that for galaxies less massive than $\sim\!3\!\times\!10^{11}\,M_\odot$ {\it in situ} star formation contributes $\sim\!50\%$ of the nucleus mass and becomes more important for more massive galaxies.~This suggests that massive galaxies are more efficient in driving the gas flows to the galaxy core regions than are low-mass galaxies.~This gas funneling allows for subsequent star formation to progress to more advanced stages with implications for the resulting chemical makeup of the stellar populations, which would exhibit lower [$\alpha$/Fe] element ratios.

In any case, the \cite{Antonini15} models require modification in order to reproduce the sharp upturn of the observed $\eta_n\!-\!{\cal M}_{\rm gal}$ relation.

\subsection{Correlation of the nucleation strength with the spatial distribution in Fornax}
Figure~\ref{fig:spatial_mr} shows the spatial distribution of the galaxies in the central region of the Fornax cluster, where the nucleated galaxies from the NGFS and ACSFCS sample with mass ratio information are shown in orange symbols, while the rest of the dwarf galaxies are indicated by gray symbols.~In this plot, the transparency of the symbols for the nucleated dwarfs is parameterized by the nucleus-to-galaxy mass ratio ($\eta_n$) as indicated in the legend in the top left corner.~We find that dwarfs located closer to massive galaxies have mass ratios below 1\%.~However, we see that the dwarfs with the highest mass ratios in our sample are located on the edges of the field of view at North and South direction.

In Section~\ref{txt:spprops} we introduced two nucleated dwarf groups according to their stellar population parameters derived from SSP model predictions in the $u'i'Ks$ diagram.~We defined group ${\cal B}$ of nuclei that appear younger and more metal-rich than the nuclei of group ${\cal A}$, which appear, on average, older and more metal-poor.~The nuclei from group ${\cal B}$ are marked with purple ellipses in Figure~\ref{fig:spatial_mr}, which reveals a strong asymmetry in their spatial distribution, where virtually all of the younger and metal-rich nuclei are located in an overdensity westwards of the Fornax center.~This suggests that dwarf galaxies in different regions of the Fornax cluster must have experienced different formation histories.~We will test the significance of this overdensity once the dwarf sample from the entire NGFS footprint is available.

\section{Conclusions}
We have characterized 61 nuclear star clusters in the Fornax cluster region ($\leq\!R_{\rm vir}/4$).~We used deep and homogeneous $u'g'i'JK_s$ photometry to obtain information on their luminosity and color distributions.~In the following we summarize our main results.
\begin{itemize}
\item The nucleation fraction ($f_{\rm nuc}$) depends strongly on the galaxy luminosity, reaching $f_{\rm nuc}\!\ge\!90\%$ for the bright NGFS dwarf galaxies ($M_{g'}\!\le\!-16$\,mag) and dropping to zero at absolute galaxy magnitudes fainter than $M_{g'}\!\simeq\!-10$\,mag.~The galaxy luminosity at which the nucleation stops corresponds to a stellar mass of $\log {\cal M}_*\!\simeq\!6.4\,M_\odot$.~As the NGFS data have a very faint point source detection limits ($M_{g'}\!\approx\!-5.4$\,mag) this is an astrophysical effect and clearly not related to observational constraints. 
 
\item Color-magnitude diagrams using various filter combinations show that nuclei occupy the bluest parts in color space, but have a comparable luminosity coverage to the distribution of compact stellar systems (CSSs) in Fornax.~The latter distribution is significantly broader, which is mainly due to the large spread in metal content.~Nuclei in dwarf galaxies show a flat color-magnitude relation, which is opposite to the trend found for UCDs and the dwarf galaxy spheroids which both show a positive color-luminosity relation. 
 
\item We derive stellar masses for our nuclei with a mean uncertainty of $\sim\!19\%$ and find that the nucleus stellar mass distribution covers the range of $\log({\cal M}_*/M_\odot)\!=\!4.8\!-\!7.3$.~We find that the nucleus mass distribution is bimodal, with peaks located at $\log({\cal M}_*/M_\odot)\!\simeq\!5.38$ and $6.25$.~The second peak is consistent with UCD masses.~We derive stellar masses for our CSS comparison dataset, which is limited at the low-mass end by the spectroscopic selection function.   
 
\item The combination of the $u'i'K_s$ diagram with SSP model predictions reveals a bimodality in the stellar population parameters of nuclei, which is congruent with the two groups in the mass distribution function of NGFS nuclei.~We define two groups with group $\cal A$ comprising nuclei with colors $(u'\!-\!i')_0\!<\!2.0$\,mag and $(i'\!-\!K_s)_0\!<\!-0.2$\,mag, which according to SSP models is consistent with metal-poor stellar populations ($Z\!<\!0.02\,Z_\odot$) and ages older than 2\,Gyr.~The nuclei in group $\cal A$ have stellar masses $>\!10^6\,M_\odot$.~Group $\cal B$ contains less massive objects and covers a more extended region in the $u'i'K_s$ color space with redder average colors, an indication of a larger range in metallicity $0.2\!<\!Z/Z_\odot\!<\!1$ and ages younger than 2\,Gyr.~With the exception of one object the masses of the group $\cal B$ nuclei are all $<\!10^6\,M_\odot$.
 
\item Dividing the CSS sample confirmed by radial velocity into GCs and UCDs using a stellar mass cut at $\log({\cal M}_*/M_\odot)=6.3$ shows that the low-mass mode of our sample nuclei (group $\cal B$) cannot be progenitors of Fornax UCDs.~On the other hand, the high-mass mode nuclei located in bright galaxies could be potential progenitors of metal-poor UCDs.~Notwithstanding, our NGFS nuclei could all be considered as potential future GCs, once their host galaxy spheroids are stripped due to the dynamical evolution of the system inside the Fornax cluster environment. 
 
\item Color differences between the nucleus and its parent galaxy spheroid correlate more with the spheroid light than with the nucleus luminosity.~Therefore, the mechanism that produces these color offsets is more likely to be acting on galaxy scales.~Colors with a wide SED coverage, such as $\Delta (u'\!-\!i')_0$ and $\Delta (u'\!-\!g')_0$, are more sensitive to changes in stellar populations and show a steeper relation with spheroid luminosity than $\Delta (g'\!-\!i')_0$.~Bright nuclei tend to be bluer than their host galaxy.~Nonetheless, as we sample fainter galaxy luminosities, we find a transition point where the color offset becomes more stochastic and we find both bluer and redder nuclei than their host.~This transition occurs at $M_{g',{\rm nuc}}\!\simeq\!-7.5$ or $M_{g',{\rm sph}}\!\simeq\!-12.5$ for $\Delta (g'\!-\!i')_0$ and $M_{g',{\rm nuc}}\!\simeq\!-9.5$ or $M_{g',{\rm sph}}\!\simeq\!-14.0$ for $\Delta (u'\!-\!i')_0$ and $\Delta (u'\!-\!g')_0$.
 
\item Scaling relations such as the nucleus-to-galaxy mass relation (${\cal M}_{\rm nuc}$ vs.~${\cal M}_{\rm gal}$) show a clear mass correlation between nuclei and their host galaxy over the entire mass range of our NGFS sample.~This relation shows a break in the slope at $\log({\cal M}_*/M_\odot)\!\simeq\!9.7$ where we find a shallower slope for dwarf galaxies relative to their more massive counterparts.~Comparing with the BH-galaxy mass relation, we find that it has a similar relation to the nuclei and their host galaxy mass down to $10^9\,M_\odot$.~For galaxies with lower masses, their nucleus-galaxy mass scaling relation becomes flatter than the BH-galaxy mass relation.~For the nuclei-to-galaxy mass ratio vs.~galaxy mass relation ($\eta_n\!=\!{\cal M}_{\rm nucleus}/{\cal M}_{\rm galaxy}\,{\rm vs.}\,{\cal M}_{\rm galaxy}$) an interesting anti-correlation is found.~The lower the galaxy mass the more prominent becomes the nucleus with a scaling $\eta_n\!\propto\!{\cal M}_{\rm gal}^{-0.5}$.~For masses higher than the break at $\log({\cal M}_*/M_\odot)\!\simeq\!9.7$ we find a positive correlation of the form $\eta_n\!\propto\!{\cal M}_{\rm gal}^{4}$.~These relatively strong trends for low-mass and high-mass galaxies reach values up to $\eta_n\!\simeq\!10\%$ for dwarf galaxies with a stellar mass of $10^7\,M_\odot$ and massive ellipticals at $10^{11}\,M_\odot$.The low-mass anti-correlation seems to be similar for ETGs and LTGs, suggesting that it is independent of galaxy type.  
 
\item The spatial distribution of the Fornax nucleated dwarfs shows that they are preferentially distributed along the East-West direction.~Knowing the location of the nucleated dwarfs with highest $\eta_n$ values, we observe that they lie at the edges of the central NGFS footprint to the North and South.~We also find that the nuclei that are members of group $\cal B$ that are relatively metal-rich and have ages younger than 2\,Gyr lie predominantly westward of NGC\,1399, suggesting a more extended star formation history of nuclei in that direction.
\end{itemize}

Our NGFS study has extended the galaxy nucleus research towards the faint galaxy luminosity regime down to $\log({\cal M}_*/M_\odot)\!\simeq\!6$, finding nucleus-galaxy scaling relations that are quite different compared to the results obtained from bright galaxies.~Theoretical models still fail to explain the observed scaling relations for the low-mass regime and do not account for the apparent transition between low-mass and high-mass galaxies.~However, we find that the models by \cite{Bekki07} and \cite{Antonini15} appear to frame our observations, which may indicate that a combination of their prescriptions may best represent reality.~Overall our NGFS nucleus sample gives crucial insights into the formation mechanism at work, showing that nuclei are likely formed via two different mechanisms, i.e.,~formation via dynamical friction acting on GCs sinking to the center and star formation processes in the central regions.~The full NGFS footprint will provide a larger sample and help us to better understand the fascinating properties and the formation mechanisms of the nucleus population in dwarf galaxies.

\acknowledgments

This project is supported by FONDECYT Regular Project No.~1161817 and the BASAL Center for Astrophysics and Associated Technologies (PFB-06).~Y.O.-B.\ acknowledges financial support through CONICYT-Chile (grant CONICYT-PCHA/Doctorado Nacional No.~2014-21140651) and the DAAD through the PUC-HD Graduate Exchange Fellowship.~T.H.P.~and A.L.~gratefully acknowledge ECOS-Sud/CONICYT project C15U02.~P.E.~acknowledges support from the Chinese Academy of Sciences (CAS) through CAS-CONICYT Postdoctoral Fellowship CAS150023 administered by the CAS South America Center for Astronomy (CASSACA) in Santiago, Chile.~M.A.T. is supported by the Gemini Observatory, which is operated by the Association of Universities for Research in Astronomy.

This project used data obtained with the Dark Energy Camera (DECam), which  was constructed by the Dark Energy Survey (DES) collaboration.~This research has made use of the NASA Astrophysics Data System Bibliographic Services, the NASA Extragalactic Database, the SIMBAD database, operated at CDS, Strasbourg, France \citep{wen00}.~This research has made use of "Aladin Sky Atlas" \citep{bon00, boc14}, developed at CDS, Strasbourg Observatory, France.~Software used in the analysis includes, the {\sc Python/NumPy} v.1.11.2 and {\sc Python/Scipy} v0.17.1 \citep[][\url{http://www.scipy.org/}]{jon01, van12}, {\sc Python/astropy} \citep[v1.2.1;][\url{http://www.astropy.org/}]{ast13}, {\sc Python/matplotlib} \citep[v2.0.0;][\url{http://matplotlib.org/}]{hun07}, {\sc Python/scikit-learn} \citep[v0.17.1;][\url{http://scikit-learn.org/stable/}]{ped12} packages.~This research made use of ds9, a tool for data visualization supported by the Chandra X-ray Science Center (CXC) and the High Energy Astrophysics Science Archive Center (HEASARC) with support from the JWST Mission office at the Space Telescope Science Institute for 3D visualization.\\

\facility{CTIO (4m Blanco/DECam)}.

\end{document}